\documentclass[preprint,authoryear,12pt]{elsarticle}

\usepackage{epsfig}

\usepackage{amssymb}

\usepackage[a4paper=true,%
breaklinks=true,%
colorlinks=true,%
pdfauthor={First Author et al.},%
pdftitle={Template for manuscripts in Advances in Space Research}%
]{hyperref}

\journal{Advances in Space Research}

\begin{document}

%%%%%%%%%%%%%%%%%%%%%%%%%%%%%%%%%%%%%%%%%%%%%%%%%%%%%%%%%%%%%%%%%%%%%%%%%%%%%
%% Frontmatter
\begin{frontmatter}


% Use the tnoteref command within \title and fnref within \author or \address for footnotes;
% use the corref command within \author for corresponding author footnotes;
% \author{Name\corref{cor1}\fnref{label2}}
% \address{Address\fnref{label3}}

%\title{Template for manuscripts in Advances in Space Research\tnoteref{footnote1}}
\title{Pulsed Laser Interactions with Space Debris: Target Shape Effects}
%\tnotetext[footnote1]{This template can be used for all publications in Advances in Space Research.}

% Use optional labels to link authors explicitly to addresses:

%\author{D. A. Liedahl\corref{cor}, A. Rubenchik, S. B. Libby, S. Nikolaev}
\author{D. A. Liedahl, A. Rubenchik, S. B. Libby, S. Nikolaev}
%\author{D. A. Liedahl\corref{cor}\fnref{footnote2}}
\address{Lawrence Livermore National Laboratory}
%\cortext[cor]{Corresponding author}
%\fntext[footnote2]{Additional information regarding the corresponding author}
\ead{liedahl1@llnl.gov,rubenchik1@llnl.gov,libby1@llnl.gov,nikolaev2@llnl.gov}

% Url can be given like this:
% \ead[url]{http://www.elsevier.com/wps/find/authorsview.authors/latex}

%\author{A. Rubenchik, S. B. Libby}
%\author{A. Rubenchik, S. B. Libby\fnref{footnote3}}
%\address{Lawrence Livermore National Laboratory}
%\fntext[footnote3]{Additional information about the second and third authors}
%\ead{rubenchik1@llnl.gov, libby1@llnl.gov}

\author{C. R. Phipps}
%\author{C. R. Phipps\fnref{footnote4}}
\address{Photonic Associates, LLC}
%\fntext[footnote4]{Additional information about the co-authors}
\ead{crphipps@photonicassociates.com}

\begin{abstract}
Among the approaches to the proposed mitigation and remediation
of the space debris problem is the de-orbiting of objects in low Earth orbit
through irradiation by ground-based high-intensity pulsed lasers. Laser ablation
of a thin surface layer causes target recoil, resulting in the depletion of orbital angular momentum
and accelerated atmospheric re-entry. However, both the magnitude and
direction of the recoil
are shape dependent, a feature of the laser-based remediation concept that has received
little attention. Since the development of a predictive capability is desirable,
we have investigated the dynamical response to ablation of objects comprising a variety of shapes.
We derive and demonstrate a simple analytical technique for
calculating the ablation-driven transfer of linear momentum,
emphasizing cases for which the recoil is not exclusively parallel to the incident beam. 
For the purposes of comparison and contrast, 
we examine one case of momentum transfer in the low-intensity regime, where photon pressure
is the dominant momentum transfer mechanism, showing that shape and orientation effects
influence the target response in a similar, but not identical, manner.
We address the related problem of target spin and, by way of a few simple examples, show how
ablation can alter the spin state of a target, which often has a pronounced effect on the recoil dynamics.
\end{abstract}

\begin{keyword}
%first keyword \sep second keyword \sep more keywords
laser ablation; laser orbit modification
\end{keyword}

\end{frontmatter}

\parindent=0.5 cm

%%%%%%%%%%%%%%%%%%%%%%%%%%%%%%%%%%%%%%%%%%%%%%%%%%%%%%%%%%%%%%%%%%%%%%%%%%%%%
%% Main text

\section{Introduction}

Over 90\% of all objects in LEO represent space
debris, including non-operational spacecraft, rocket stages, mission-related
debris, fragmentation debris, NaK coolant droplets, Al$_2$O$_3$ slag from solid
rocket motor firings, etc. \citep{klinkrad}.
High-velocity impacts of operational payloads with debris objects as small as
1 cm in size and a few grams in mass can have potentially devastating effects:
the on-orbit collisions produce more debris that increase the risk of
subsequent orbital collisions, leading to a potential self-sustaining, runaway
debris-creating process, known as the Kessler syndrome \citep{kessler78,kessler89}.
To counter the threat posed by the Kessler syndrome, several debris mitigation
strategies have been proposed, including post-mission disposal of spacecraft,
passivation of upper rocket stages, maneuvering into graveyard orbits for
geostationary satellites, and improved collision avoidance.

Remediation (as distinct from mitigation) of the debris threat has also been proposed, and was given 
consideration under the auspices of NASA's Orion Project in the
1990s \citep{campbell96}, in which 
a candidate remediation concept called for de-orbiting the debris using 
a ground-based laser \citep{phipps96,phipps98}. Predating the Orion Project, a somewhat more exotic
technique involving a laser-mounted autonomous orbital vehicle had
been suggested \citep{schall91,schall98}. In either case, the key concept is the creation
of an ablation jet on the debris fragment, whereby orbital angular momentum can be
removed, lowering the perigee and accelerating re-entry.

The Orion Project grouped the debris into five ÒlikelyÓ compositional classes: aluminum, steel, 
NaK metal, carbon phenolics, and insulation. Only the debris larger than approximately 10 cm,
about 5-10\% of the total by number, can be tracked by radar and optical systems on a continuous basis.
Nevertheless, there have been efforts to garner information related to smaller debris fragments; 
surveys in the radio and optical bands, augmented with modeling \citep{sdunnus}, suggest that the population 
of objects in the 1--10 cm size range is $>150,000$, each member of which is too small to be monitored 
continuously, but any of which, with impact velocities of up to 16 km s$^{-1}$, 
can debilitate or destroy spacecraft functionality. Since damage 
from collisions with objects smaller than about 1 cm can be mitigated against by shielding, it is 
the group of objects in the 1--10 cm size range that pose the greatest risk to space assets, 
but which, owing to their relatively small masses, can be favorably maneuvered to lower orbits
with short re-entry times. Deflections of larger objects for the purpose of 
collision avoidance are also feasible \citep{phipps2012}.

The current microscopic picture of laser debris coupling has its origin in studies of 
direct-drive physics relevant to inertial confinement fusion --- the picture developed 
(independently) by Kidder, Caruso, and Nemchimov \citep{kidder}. 
For practical applications, the connection between the microphysics and the macrophysics
is made by way of the mechanical coupling coefficient, denoted 
here by $C_m$. For a given element of surface of mass $m$ oriented such that its
surface normal vector is labeled by $\hat{n}$, $C_m$ is defined so that the momentum change 
of a target responding to a laser energy deposition is given by
\begin{equation}
\label{eq:basic}
m \Delta \vec{v} = -C_m E_{\rm inc} \, \hat{n},
\end{equation}
where $E_{\rm inc}$ is the total on-target laser energy. 
This expression contains the presumption that the net momentum vector of the ablation flow
is parallel to the local surface normal. 

Given Eq.\  \ref{eq:basic}, there is no reason to expect that
momentum is transferred strictly along the laser propagation vector $\hat{k}$. However, that assumption is often used
in estimating target recoil and subsequent orbital modifications, and constitutes one of the underpinnings of concept studies 
of debris clearing with lasers.
The assumption that $\Delta \vec{v}$ and $\hat{k}$ are parallel simplifies calculations of the 
orbital dynamics and establishes a prescription for removing orbital angular momentum 
from the target: uprange irradiation, as close as possible to the horizon, after due consideration 
is given to geometrical dilution and atmospheric effects on the beam. 

It is true that some objects {\it do} recoil in the beam direction; spheres, for example.
Indeed, the concept of laser remediation, including the quantitative characterization 
of engagement strategies, is founded largely upon the assumption of spherical 
targets. While some fraction of the space debris population is likely to be spherical
(see below), the actual shape distribution is uncertain, since 
evaluations of debris characteristics based on radar data 
do {\it not} include reliable shape assessments. Still, according to 
the NASA Technical Memorandum 108522 \citep{orion}, periodicities of 
debris signatures have been noted, with periods ranging from about  0.1 second to 
tens of seconds. This suggests that at least some fraction of the debris is 
non-spherical or, at least, irregular. In any case, in light of Eq.\ 1, it is clear that
a more general treatment of the problem is in order.

Given a differential element of area with a specific orientation characterized
by its surface normal vector,
the mechanical response following interaction with the laser is tied to 
(1) the orientation of the surface normal relative
to the position vector with respect to Earth's center, which, in part, 
determines the angular momentum change to the orbit, and (2) the 
orientation of the surface normal with respect to the laser propagation vector, 
which determines the incident laser fluence and the impulse. We can anticipate a variety of possibilities: 
objects consisting essentially of two faces, such as a thin metallic plate, objects consisting 
of several faces, such as a rectangular solid, debris characterized by a continuously varying 
surface normal, such as a cylindrical or needle-like object, or irregularly-shaped objects 
that defy simple descriptions. In all these cases, the orbital solutions differ from the case of a 
spherical target.

Other laser-based schemes for space debris remediation or maneuvering must also accommodate
shape effects, although the specifics can vary. For example, if, rather than surface ablation, photon pressure
is the mechanism through which momentum is imparted to a target  \citep{mason}, then the recoil dynamics depend on
the relative amounts of absorption, diffuse reflection, and specular reflection, the first
transmitting momentum along the beam, and the other two adding off-beam components that are
anti-parallel to the surface normal. 
\citet{mason} argue that target spin will tend to average the specular
reflection component to zero, and do not treat off-beam momentum transfer. 
However, we show below, explicitly for the ablation case, that the momentum impulse transmitted
to a spinning object often retains a component
transverse to the beam, and that the magnitude of this component is sensitive to the
initial condition, i.e., the orientation of the target at the onset of laser illumination.

We have performed some preliminary work on the topic of shape effects \citep{liedahl}, and
expand upon it here. Our focus in this paper is on shape effects in the target rest frame. The 
consequences relevant to orbital modifications will be addressed in a followup paper. 
In \S2, we develop the physical context of the problem and discuss some of the
assumptions used in this paper. The basic approach for dealing with linear
momentum transfer for arbitrary shapes is presented in \S3. We present the results of
a variety of calculations  based on this methodology in \S4. In \S5, the related problem
of target spin is addressed, again featuring a few examples. We conclude in \S6,
with a summary of our results, and suggestions for future refinements.

\section{A Few Preliminaries}

To better define the context of the problem, we derive some approximate scaling relations
and discuss an approximation used throughout the paper.

\subsection{Estimate of the Required Velocity Change}

To obtain an estimate of the required velocity change, we consider the Hohmann transfer
\citep{bate} from a circular orbit with radius $r$ to an elliptical orbit with perigee $r_p$, as shown in Fig.\ 1.
From the the so-called {\it vis viva} equation, the orbital velocity of an object on an elliptical
orbit around a mass $M$, taken here to be the mass of Earth, with semi-major axis $a$ is given by
\begin{equation}
\label{eq:visviva}
v^2=\frac{2GM}{R}-\frac{GM}{a},
\end{equation}
when it is a distance $R$ from Earth's center. At the ``transfer point,'' we thus have
\begin{equation}
v_1^2=\frac{GM}{r} ~~~~ v_2^2=\frac{2GM}{r}-\frac{2GM}{r+r_p},
\end{equation}
where the ``1'' and ``2'' denote ``before (circle)'' and ``after (ellipse),'' respectively. If $\Delta v= v_2 - v_1$, then
\begin{equation}
\frac{\Delta v}{v_1} =\biggl(\frac{2r_p}{r+r_p} \biggr)^{1/2} -1.
\end{equation}
Considering LEO only, and letting $\Delta r =r_p-r$, an adequate approximation is
\begin{equation}
\label{eq:dvreq}
\Delta v \approx v_1 ~ \frac{ \Delta r }{4R_E} \approx
30 ~\frac{ \Delta r}{100~{\rm km}} ~~{\rm m ~s^{-1}},
\end{equation}
where $R_E$ is Earth's radius, and we 
set  $v_1$ to 8000 m s$^{-1}$. Therefore, LEO debris clearing
requires velocity changes of up to a few hundred m s$^{-1}$, with the required $\Delta v$
directly (though approximately) proportional to the desired $\Delta r$. Note that it is also
feasible to boost debris such that they achieve escape velocity. While perhaps
not the best choice, since it merely displaces the problem, albeit into a much larger
volume, it is in any case energetically expensive, since the required $\Delta v$ (in this case, positive)
is approximately $\sqrt{2}-1$ times the orbital velocity, or about 3000 m s$^{-1}$ for LEO.

\begin{figure}
\begin{center}
\includegraphics*[width=7cm,angle=0]{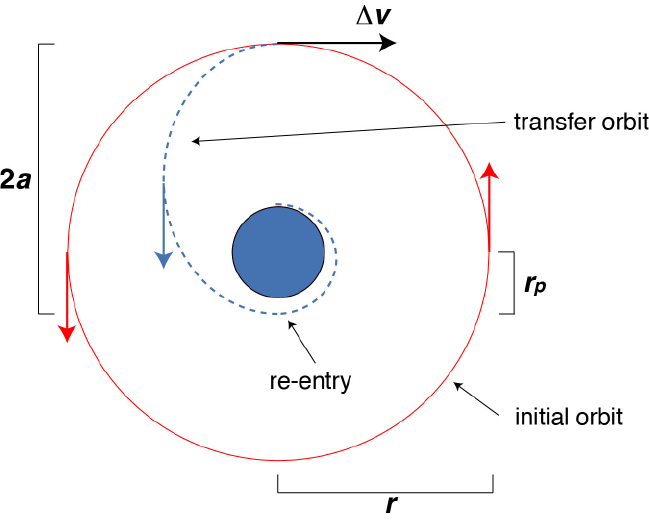}
\end{center}
\caption{Geometry used to estimate the velocity change required to obtain a pre-determined perigee
{\it via} the initial stage of a Hohmann transfer (scales exaggerated for clarity).
Starting from a counter-clockwise circular orbit ({\it solid curve}) with radius $r$,
an impulse opposite the velocity vector, resulting in a velocity change $\Delta v$, 
is applied at the transfer point. The perturbed orbit (elliptical until capture; {\it dashed curve}) is characterized by
its semi-major axis $a$ and perigee $r_p$, where $r_p$ is sufficiently low for re-entry to occur.}
\end{figure}

The estimate given above is just that, provided primarily to establish the
regime in which we are working, and to serve as a useful benchmark for 
estimating laser energy requirements. Of course, one cannot simply 
choose to effect a classic Hohmann transfer from a ground
station, since the momentum impulse given to an orbiting object will (typically) 
have components transverse to the velocity vector. We note the extreme case of a 
pure radial impulse, which does not remove {\it any} angular momentum, 
and, in fact, {\it adds} energy, but which nevertheless leads to a perigee reduction. 
Moreover, space debris orbits are not precisely circular. Considerations
bearing on the efficient use of laser energy must make a distinction between 
engagement at descent and engagement at ascent. The various relations among 
orbital parameters, ground station coordinates, and target characteristics lead 
to a vast range of possibilities. Our interest here is to begin to characterize
that range of possibilities, emphasizing the importance of target shape. 

\subsection{Mass Loss}

The well-known rocket equation gives the velocity change of an object that has undergone continuous
mass loss at a constant rate in a directed manner,
with a mass ejection velocity $v_a$ in the rest frame of the object whose initial mass is $m_o$;
\begin{equation}
\label{eq:rocket}
\Delta v = v_a \ln (1-\Delta m/m_o)^{-1} \approx \frac{\Delta m}{m_o} \, v_a,
\end{equation}
where the approximation is valid when the fractional mass loss is small.
As is customary, in the model calculations presented here, we disregard the dynamical 
effects of material removal for all cases of linear momentum transfer. 

Taken in its simplest form, active perigee reduction requires a change to the debris velocity $\Delta v$
as given in the previous subsection. Combining the approximation in Eq.\  \ref{eq:rocket} with the
expression for $\Delta v$ from Eq.\  \ref{eq:dvreq}, we can estimate the fractional mass loss by
\begin{equation}
\frac{\Delta m}{m}  \approx \frac{v_1}{v_a} ~ \frac{ \Delta r }{4R_E}.
\end{equation}
The ejecta velocity $v_a$ depends on the material; it is higher for low-Z material than for heavy elements, and is 
about the thermal velocity of the surface plasma. To take an example, for Al $v_a$ is about 3-5 km s$^{-1}$. Therefore,
as an approximation, the fractional mass loss is
\begin{equation}
\frac{\Delta m}{m}  \approx
10^{-2} ~\frac{ \Delta r}{100~{\rm km}}.
\end{equation}
Thus we can anticipate a total mass loss of order 1--10\% of the initial mass
for LEO de-orbiting campaigns. The fractional error introduced by using an approximate
$\Delta v$, rather than the exact value $\Delta v_{\rm rocket}$ is, from
Eq.\ \ref{eq:rocket},
\begin{equation}
\frac{\Delta v-\Delta v_{\rm rocket}}{\Delta v_{\rm rocket}}=-\frac{1}{2} \, \frac{\Delta m}{m_o} 
+ O \biggl(\frac{\Delta m}{m_o} \biggr)^2.
\end{equation}

The ablated material expands, cools down, and forms nanoclusters 
\citep{zeldovich}, which are harmless. It is not clear, however, that ignoring shape deformation
is ``harmless'' in terms of developing a predictive capability.  Nevertheless,
in the interest of keeping the calculations analytically manageable, we have not generally
accounted for shape evolution during the course of an engagement. 

We postpone our discussion of the treatment of mass loss in the context of angular momentum
transfer until \S5.1, after we develop the relevant analog of the force equation in the rotational case. 

\subsection{Approximating the Coupling Coefficient}

Given that the emphasis of the paper is on shape effects, the {\it magnitude} of the
coupling coefficient $C_m$ is not crucial. However, since the magnitude is known to depend
on the incident intensity, and since we expect the intensity to vary across a non-planar surface,
we need to justify an approximation used throughout the paper, namely, that $C_m$ can be
treated as a constant across the illuminated surface of a given target.

\begin{figure}
\begin{center}
\includegraphics*[width=13cm,angle=0]{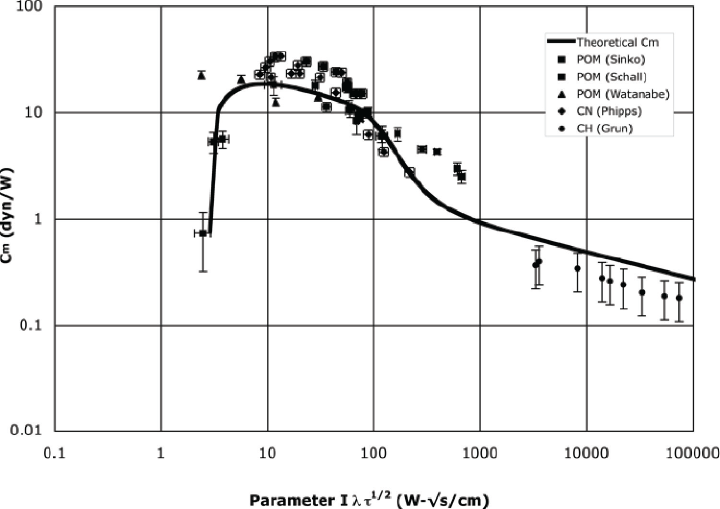}
\end{center}
\caption{The mechanical coupling coefficient for several materials
(cellulose nitrate, polyoxymethylene, polystyrene) plotted as a function
of $I \lambda \tau^{1/2}$, for intensity $I$, wavelength $\lambda$, and pulse duration
$\tau$, illustrating the broad, slowly
varying plateau in the region of maximum coupling, at intensities just above those
that initiate plasma formation. The laser intensity corresponding to 10
on the $I \lambda \tau^{1/2}$ scale, near the peak in $C_m$,
is $10^9$ W cm$^{-2}$ for a ns pulse at laser wavelength of 1 $\mu$, the value we
assume here for numerical estimates. Figure from \citet{sinko}.}
\end{figure}

A review of the experimental data for $C_m$ is presented by \citet{phipps88}, and typical 
data are shown in Fig.\ 2 \citep{sinko}. 
As a function of laser intensity, $C_m$ is peaked not far beyond 
the vaporization threshold, where plasma starts to be generated. At low intensity levels, the surface
temperature and evaporation rate are low, and the recoil momentum is relatively small ---
the coupling increases with intensity in this regime. At high 
intensity levels, a large fraction of the laser energy is used to create a plasma, which contributes
little to the momentum change of the debris --- the coupling decreases with intensity in this regime. Therefore,
for some critical value of the laser intensity $I_{\rm crit}$, $C_m$ is maximized.
But, as is evident from Fig.\ 2, the maximum does not sharply define $I_{\rm crit}$; $C_m$
achieves a plateau over a broad range of laser intensity. 

Since it is advantageous to work in the regime of maximum coupling,
laser systems designed for ICF applications are well-suited to space debris clearing,
as discussed in \citet{sasha12}. In fact, data from different groups 
demonstrates that for a broad range of wavelength, pulse duration, and pulse energy, 
the coupling coefficient maximum is reached at an intensity
\begin{equation}
\label{eq:icrit}
I_{\rm crit}=\frac{2.5}{\tau^{1/2}} ~~{\rm GW~cm^{-2}},
\end{equation}
where $\tau$ is the pulse duration in nanoseconds \citep{phipps88,sasha12}. 
This points to high-powered short-pulse lasers. For intensities near $I_{\rm crit}$ typical values of 
$C_m$ are 1--10 dyne W$^{-1}$, and over the wide range of intensities near the peak, 
$C_m$ can be treated as being roughly constant. We adopt this approach below. 

The experimental data correspond to planar targets only. It is fair to wonder if these data
are applicable to calculations emphasizing non-planar targets. We note that for nanosecond-scale pulses the energy 
is absorbed in a thin surface layer, and the material is ejected normal to the surface. 
For ejecta velocities of about a few km s$^{-1}$, the ablation jet extends only a few microns during the pulse. 
Locally, the situation is equivalent to a flat surface.

\subsection{Estimate of the Required Laser Energy}

To relate $E_{\rm inc}$ from Eq.\ \ref{eq:basic} to the corresponding laser energy at the site of generation, the
spot size is the crucial factor. In terms of the beam quality $Q$, the laser wavelength $\lambda$,
the distance of the target from the laser site $d$, and the diameter of the beam director $D$, an
expression for the spot size $a$ that accounts for beam diffraction is \citep{siegman}
\begin{equation}
\label{eq:spotsize}
a= \frac{2 \lambda \, Q^2 d}{\pi D}.
\end{equation}
As discussed by \citet{mason}, to approach the diffraction limit, a de-orbiting 
laser system would require the use of adaptive optics techniques with an artificial guide star,
so as to mitigate against beam divergence caused by, for example, atmospheric turbulence \citep{beckers}. 
Then, for a target of cross-sectional area $A$ as projected onto the beam direction,
we have from Eq.\  \ref{eq:basic}
\begin{equation}
\label{eq:elaser}
E_{\rm inc} =\frac{E_L A}{\pi a^2}=\frac{m \Delta v}{C_m}.
\end{equation}
Assuming that the area-to-mass ratio is about $(\rho^2 m)^{-1/3}$, 
we find from Eqs.\ \ref{eq:dvreq}, \ref{eq:spotsize}, and  \ref{eq:elaser} that the required laser energy 
at the source can be expressed as
\begin{equation}
E_L=\frac{(\rho^2 m)^{1/3}}{\pi C_m} \, \frac{\lambda^2 \, Q^4}{D^2} ~ d^2 v ~ \frac{\Delta r}{R_E}.
\end{equation}
In order to provide a numerical estimate, we take a 100 g target, with $C_m=10$, 
$\lambda=1 \,  \mu$, $Q^2=2$, and scale to the remaining parameters, which gives
\begin{equation}
\label{eq:ereq}
E_L \sim 10^3 ~ \biggl( \frac{D}{6 ~{\rm m}} \biggr)^{-2} \, \biggl( \frac{d}{500 ~{\rm km}} \biggr)^{2}
\, \biggl( \frac{\Delta r}{100 ~{\rm km}} \biggr) ~ {\rm kJ}.
\end{equation}
With a high rep-rate laser facility, such as those envisioned for future ICF
power plants \citep{norimatsu,dunne}, operating at, say, 10 Hz, with a 10 kJ per few-ns pulse, 
approximately $10^4$ kJ can be generated on a timescale comparable to a single-pass 
engagement (a few minutes), which compares favorably with Eq.\ \ref{eq:ereq}, even if one
chooses a less optimistic $C_m=1$.

\section{Transfer of Linear Momentum}

As discussed earlier, it is not valid to assume that the direction of
the impulse following laser engagement is parallel to the beam, since that assumption is counter
to the claim that ablation proceeds in a direction parallel to the surface normal.
In this section we expand upon Eq.\  \ref{eq:basic}, and develop a methodology that 
will serve as a foundation for dealing with the general case of those irregularly-shaped
objects that are likely to constitute a large fraction of the space debris population. While
we deal strictly with idealized shapes, it is our aim to investigate some fundamental aspects
of the problem that will provide some insight and form the basis of more advanced numerical studies.

Another case for which off-beam momentum transfer is important is that arising
from interactions with targets that are larger than the beam itself. We do not treat this
type of interaction in this paper, but rather assume throughout that the beam fully envelops the target.
With beam radii of a few tens of cm (see Eq.\ \ref{eq:spotsize}), this is not particularly restrictive, since
we are interested here in debris objects in the 1-10 cm range. Moreover, we ignore spatial variations
in the beam intensity over the length scale of a debris surface, since, in the case of small debris, these
variations are not large. Of course, one may choose to model the interaction with a higher degree of fidelity
by including not just spatial variations of the beam (which implies a space-dependence to the
mechanical coupling coefficient)  but temporal profiles, as well. In the future, such detailed calculations
may become warranted, but are unlikely to introduce new qualitative effects that would substantially alter
the results presented here. 

Proceeding, first we recast Eq.\ \ref{eq:basic} as a force equation,
\begin{equation}
\label{eq:inst}
m \, \frac{d\vec{v}}{dt}=-C_m \, \frac{d E_{\rm inc}}{dt} \, \hat{n}.
\end{equation}
The laser intensity, as above, is denoted $I$ (energy per unit area per unit time).
More rigorously, the quantity of interest is the flux. However, for a unidirectional beam,
there is no important distinction between the scalar flux and the intensity. Nevertheless, we
preserve the vector nature of the flux here, and, following convention, denote the intensity $\vec{I}=I \hat{k}$
for a unidirectional beam with propagation unit vector $\hat{k}$. Then, for a
surface element of area $A$,
\begin{equation}
\label{eq:dedt}
\frac{d E_{\rm inc}}{dt}=-AI \hat{k} \cdot \hat{n}.
\end{equation}
The negative sign is required, since surface illumination can only occur for
surface elements such that $\hat{k} \cdot \hat{n} <0$.

After substitution from Eq.\  \ref{eq:dedt}, Eq.\  \ref{eq:inst} becomes
\begin{equation}
\label{eq:basicforce}
m \, \frac{d\vec{v}}{dt}=C_m IA \,  \hat{k} \cdot \hat{n}\hat{n},
\end{equation}
which represents the response to irradiation of a single oriented surface. In general,
to accommodate all irradiated surfaces, as well as cases where the surface normal varies
continuously on a macroscopic level, we can define the dyadic form ${\bf G}$ according to
\begin{equation}
\label{eq:gdef}
{\bf G}=\sum_{\alpha} A_{\alpha} ~ \hat{n}_{\alpha}  \hat{n}_{\alpha}
\rightarrow \int dA ~\hat{n}\hat{n},
\end{equation}
which we refer to as the {\it area matrix}.
The momentum transfer equation --- or, force equation --- can thus be written as
\begin{equation}
\label{eq:gform}
m \, \frac{d\vec{v}}{dt}=C_m I  \, \hat{k} \cdot {\bf G}.
\end{equation}
A matrix element $G_{ij}$ (note that $G_{ij}=G_{ji}$) describes the momentum impulse along the $j$-th coordinate direction
resulting from the $i$-th component of the laser intensity vector.
This could be the starting point for deriving a modified version of the rocket equation,
but since we are working in the limit that the mass loss is small compared to the initial mass,
we will accept Eq.\ \ref{eq:gform} with $m$ constant, as implied.
It should also be noted that ${\bf G}$ may be time-dependent; if any or all of the surface normals
change with respect to a fixed coordinate system, then ${\bf G} \rightarrow {\bf G}(t) $.

\subsection{Approximating the Pulse Train}

We refer to a set of $N$ laser pulses with $N$ ablation events as an
{\it engagement}. Roughly speaking, we assume that an engagement has a duration of
tens of seconds, comprising some hundreds of individual pulses.
Since we are considering nanosecond lasers with repetition rates of order 1-10 per second,
it is a good approximation to represent the laser intensity as consisting of a set of delta functions,
with pulses occurring at $t_1$, $t_2$, ..., $t_N$. 
However, it is simpler to treat the dynamical problem
as though the intensity were constant in time throughout the engagement, in which case another
approximation --- in the other direction, as it were --- is called for.
Letting $f$ denote the laser fluence per pulse at the target position (which we treat as a
constant for a given engagement), the intensity can be written as
\begin{equation}
\label{eq:pulse}
\vec{I}= f \hat{k} \, \sum_{n=1}^N  \, \delta(t-t_n),
\end{equation}
for a series of $N$ identical pulses --- the pulse train.
Substituting this expression for the pulse train into Eq.\ \ref{eq:gform} gives
\begin{equation}
\label{eq:deltaform}
\frac{d\vec{v}}{dt}=\frac{C_m}{m} \, f \, \hat{k} \cdot  \sum_{n=1}^N  \, \delta(t-t_n) \, {\bf G}(t),
\end{equation}
where, as mentioned above, we allow for the possibility that ${\bf G}$ is time-dependent, as it may be for
a rotating object. Starting from rest,
\begin{equation}
\label{eq:steprocket}
\vec{v}(t)=\frac{C_m }{m}~f \, \hat{k} \cdot  \sum_{n=1}^N \, H(t-t_n) ~ {\bf G}(t_n),
\end{equation}
where $H$ is the Heaviside step function.

In terms of the laser repetition rate $\nu$, we make the approximation
\begin{equation}
\sum_n \, H(t-t_n) \rightarrow \nu \int dt,
\end{equation}
and define the average intensity $\bar{I}=f \nu$, so that
\begin{equation}
\label{eq:tdepg}
\vec{v}(t)=\frac{C_m}{m} ~\bar{I} \, \hat{k} \cdot  \int dt ~ {\bf G}(t),
\end{equation}
or, working backwards,
\begin{equation}
\label{eq:tdepforce}
\frac{d\vec{v}}{dt}=\frac{C_m}{m} ~\bar{I} \, \hat{k} \cdot  {\bf G}(t),
\end{equation}
where we recover Eq.\ \ref{eq:gform} (now generalized).
Problems involving both time-independent and time-dependent variants
of ${\bf G}$ are worked out below using Eq.\ \ref{eq:tdepforce}. 

For the remainder of the paper, we replace the symbol $\bar{I}$ by $I$, with the understanding that $I$ 
is the time-averaged magnitude of the laser intensity ($\bar{I}=I \nu \tau$ for a pulse duration $\tau$ of a few ns).
To get a numerical estimate of $\bar{I}$, a typical fluence $f$ at the target is about 30 J cm$^{-2}$, for
a 10 kJ pulse and a spot radius of 10 cm. If the laser pulse repetition rate $\nu$ is 10 per second, then
the time-averaged intensity $f \nu \sim 300$ W cm$^{-2}$. Where needed, we use the value 100 W cm$^{-2}$ below,
keeping in mind that this average value is used in the force equations to reproduce the dynamics of
a pulsed laser that has a much higher intensity during a pulse, so as to preserve the assumption of ablation
near the peak range of the mechanical coupling coefficient.

One can invent hypothetical cases for which the assumption of a time-constant beam
will not apply. For example, suppose a debris object is rotating such that
its period matches the laser pulse period, i.e., the object and the laser are
``phase-locked.'' In that case, only one orientation of the object is sampled.
Examples such as this can be worked out analytically. One finds that
the net velocity change at the end of the laser engagement can be less than, greater than, or even equal
to that obtained from using Eq.\ \ref{eq:tdepforce}. 
Built into the time-constant-beam approach is the assumption that ${\bf G}$ is sampled continuously,
rather than in discrete jumps. For this paper, which is intended as an exploration of some of the
basic ideas, with a few worked-out examples to illustrate those ideas, we choose to avoid
pathological cases, such as the one discussed here, and we restrict the calculations to those
involving laser repetition rates that are substantially higher than the debris rotation rates. 
Further exploration of the wide range of possibilities may constitute worthwhile subjects of future efforts.

\section{Shape Effects on Translational Motion}

To demonstrate some of the consequences of Eq.\ \ref{eq:tdepforce}, we provide a few
examples using shapes that are, admittedly, highly idealized, in that we are unlikely to
encounter many examples of such perfect objects in LEO. The examples are
intended as a first foray into some of the anticipated phenomena associated with the laser/debris
interaction that have yet to be addressed in detail.

\subsection{Cube}

The cube provides a useful example, as it demonstrates the utility of the area matrix.  In the interest
of emphasizing its relative ease of use, we first determine the momentum impulses
on a cube in two orientations without invoking the area matrix technique.
Consider a uniform cube of side $s$ and density $\rho$, so that $m=\rho s^3$.
Orient the cube so that its edges are parallel or perpendicular to the axes of a Cartesian coordinate system,
as in Fig.\ 3, so that the six surface normals coincide with the three Cartesian unit vectors
and their opposites. 

We evaluate the momentum impulse for two cases, one corresponding to the
minimum projected area $s^2$, which occurs if the illumination is anti-parallel to one of the six faces,
and the other corresponding to the case of maximum projected area $s^2 \sqrt{3}$, which occurs when the beam is
along the main diagonal of the cube, and three faces are simultaneously illuminated.
Intuition would suggest that various orientations
of the cube should result in a complex variety of 
momentum impulses. Additionally, the incident laser energy depends on the orientation; one might 
expect the maximum magnitude of the impulse
to correspond to the maximum projected area. As we now show, this is not the case.

In either case, fix the laser such that $\hat{k}=-\hat{z}$. The case of single-surface ablation requires
no rotation of the cube, whereas the down-the-diagonal case does. Note that in the context of this paper one can use
the equivalent procedure of ``moving'' the laser to access any case of interest, i.e., one can freely select $\hat{k}$. 
However, in the larger context of calculating the momentum impulse of a debris target that is on orbit, $\hat{k}$
is dictated by the laser-to-target position vector --- there is no leeway. Thus it is conceptually simpler to
rotate the object in a fixed coordinate system. We use the latter, albeit somewhat more algebraically involved, approach here. 

Referring to Fig. 3, only the top surface is illuminated in the minimum-projected-area case, 
with the surface normal $(0, 0, 1)$. Equation \ref{eq:basicforce} gives us
\begin{equation}
\label{eq:minarea}
\frac{d\vec{v}}{dt}=-\frac{C_mI s^2}{m} \, \hat{z} =\frac{C_mI s^2}{m} \, \hat{k} 
\end{equation}

\begin{figure}
\begin{center}
\includegraphics*[width=7cm,angle=0]{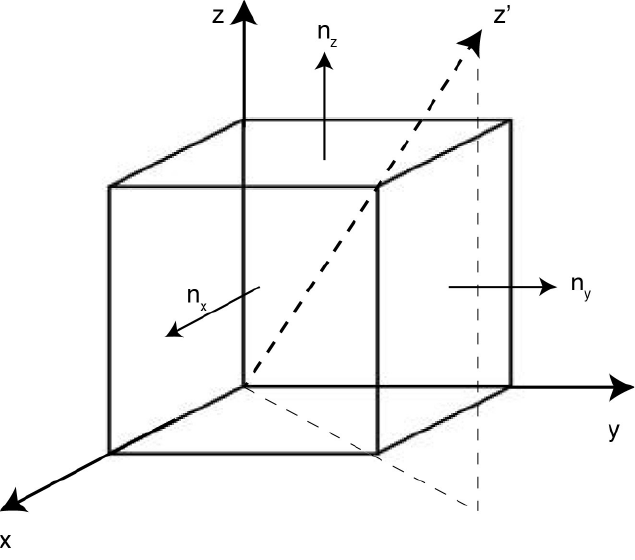}
\end{center}
\caption{Homogeneous cube of mass $m$, mass density $\rho$, and with edge length $s$, 
with three of the six surface normals illustrated. 
In one example (see text), the laser is incident from above ($\hat{k}=-\hat{z}$), so that only the 
face corresponding to $\hat{n}_z$ is illuminated. In a second example,
the cube is rotated so that the main diagonal is parallel to $\hat{k}$, i.e., $\hat{z} \rightarrow 
\hat{z}^{\prime}$, and three faces are illuminated.}
\end{figure}

For the second case, we construct a rotation matrix as the product of two coordinate rotation matrices,
the first obtained by rotation about the $z$-axis by the angle $\phi=-\pi/4$ and the second by 
rotating about the $x^{\prime}$ axis (which is subsequently discarded)
by the angle $\theta=-\arccos (1/\sqrt{3})$. These are the 
first two of the three Euler angles using the ``$x$-convention'' \citep{goldstein}. Explicitly,
the net rotation proceeds as follows:
\begin{equation}
R=R_{\theta}R_{\phi}=
\frac{1}{\sqrt{3}}
\left(
\begin{array}{rrr}
\sqrt{3} & 0 & 0 \\
0           & 1             & -\sqrt{2} \\
0 & \sqrt{2}  & 1
\end{array}
\right)  \times
\frac{1}{\sqrt{2}}
\left(
\begin{array}{rrr}
1 & -1 & 0 \\
1           & 1             & 0 \\
0 & 0  & \sqrt{2}
\end{array}
\right) 
\end{equation}
which leaves
\begin{equation}
R=\frac{1}{\sqrt{6}} \left(
\begin{array}{rrr}
\sqrt{3} & -\sqrt{3} & 0 \\
1           & 1             & -2 \\
\sqrt{2} & \sqrt{2}  & \sqrt{2}
\end{array}
\right)
\end{equation}
which moves the main diagonal such that it lies along the $z$-axis.

Operating with $R$ on the three surface normals gives
\begin{equation}
\hat{n}_x^{\prime}=\frac{(\sqrt{3}, 1, \sqrt{2})}{\sqrt{6}} ~~~
\hat{n}_y^{\prime}=\frac{(-\sqrt{3}, 1, \sqrt{2})}{\sqrt{6}} ~~~
\hat{n}_z^{\prime}=\frac{(0, -2, \sqrt{2})}{\sqrt{6}} 
\end{equation}
These constitute the set of illuminated surface normals of our rotated cube in the fixed Cartesian coordinate system.

Again, using Eq.\ \ref{eq:basicforce}, the acceleration components are
\begin{equation}
\frac{dv_i}{dt}=\frac{C_m Is^2}{m} \sum_{\alpha}(\hat{k} \cdot \hat{n}_{\alpha}^{\prime})
(\hat{x}_i \cdot \hat{n}_{\alpha}^{\prime})
\end{equation}
which for this case gives the results
\begin{equation}
\frac{dv_x}{dt}=-\frac{C_m Is^2}{\sqrt{3} \, m} \times \frac{1}{\sqrt{6}} \, (\sqrt{3} -\sqrt{3}) =0
\end{equation}
\begin{equation}
\frac{dv_y}{dt}=-\frac{C_m Is^2}{\sqrt{3} \, m} \times  \frac{1}{\sqrt{6}} \, (1 +1-2)=0
\end{equation}
\begin{equation}
\frac{dv_z}{dt}=-\frac{C_m Is^2}{\sqrt{3} \, m} \times \frac{1}{\sqrt{6}} \, (\sqrt{2} + \sqrt{2} + \sqrt{2})=
-\frac{C_m Is^2}{m}
\end{equation}
The final result for the acceleration vector when illumination is down the main diagonal is thus
\begin{equation}
\label{eq:cubeforce}
\frac{d\vec{v}}{dt}= \frac{C_m Is^2}{m} \, (0, 0, -1)
=\frac{C_m Is^2}{m} \, \hat{k} =\frac{1}{(\rho^2 m)^{1/3}} ~C_m I \, \hat{k},
\end{equation}
which is identical to the case of single-surface normal-incidence illumination (Eq.\ \ref{eq:minarea}).
Moreover, as can be shown, one obtains the same result for {\it any} orientation of the cube.
This result is at odds with a scalar version of the momentum equation, since, if
that were correct, the recoil would be proportional to the area of the cube projected onto the beam
direction, which determines the amount of intercepted laser energy. Mathematically,
this can be understood as follows.
Since, in the coordinate system of Fig.\ 3, the area matrix before rotation 
is ${\bf G}= s^2 \, {\bf I}$, where $\bf{I}= \rm{diag} (1, 1, 1)$, any rotation of the object 
will leave ${\bf G}$ unchanged. For {\it any} object, if $\bf{G} \propto \bf{I}$, then
\begin{equation}
R \cdot {\bf G} \cdot R^{-1} \propto R \cdot {\bf I} \cdot R^{-1} = {\bf I},
\end{equation}
and, from Eq.\ \ref{eq:gform}, $d \vec{v}/dt \propto \hat{k} \cdot {\bf G} ~ \vert \vert ~ {\hat k}$ regardless of orientation.
Thus in general, if ${\bf G}$ takes the form of a multiple of the unit tensor in one orientation, it takes that
form in {\it all} orientations, and the recoil is along the beam. 

By contrast to the somewhat laborious calculation by which we arrived at Eq.\ \ref{eq:cubeforce},
one can use Eq.\ \ref{eq:gform} to immediately obtain the correct result
for an arbitrarily oriented cube. Qualitatively, the result expresses the
compensating trade-offs in the competition between the projected areas of the three faces and the 
transmission of momentum along the three coordinate axes (the projections of the surface normals onto
the coordinate axes), including the effects of cancellation or reinforcement. The relative convenience 
of the area matrix is evident.

\subsection{Sphere}

Since a large proportion of earlier work related to de-orbiting small debris objects has assumed
a spherical shape, we take as our next example a solid, homogeneous sphere of radius $R$
and mass density $\rho$. In fact, NaK coolant spheres are believed to constitute a non-negligible fraction of
the space debris population \citep{wiedemann}. In any case, this example serves as a straightforward application
of the area matrix approach.

Let the center of the sphere be
coincident with the origin of a Cartesian coordinate system, with the normal vector given in terms
of standard spherical coordinates according to 
\begin{equation}
\hat{n}=(\sin \theta \cos \phi, \sin \theta \sin \phi, \cos \theta),
\end{equation}
and the differential element of solid angle $d\Omega = \sin \theta \, d\theta \, d\phi$.
Using the integral form of Eq.\ \ref{eq:gdef}, the area matrix is
\begin{equation}
{\bf G}=R^2 \int d \Omega
\left(
\begin{array}{ccc}
\sin^2\theta \cos^2 \phi            & \sin^2 \theta \sin \phi \cos \phi & \sin \theta \cos \theta \cos \phi\\
\sin^2 \theta \sin \phi \cos \phi & \sin^2\theta \sin^2 \phi             & \sin \theta \cos \theta \sin \phi \\
\sin \theta \cos \theta \cos \phi & \sin \theta \cos \theta \sin \phi  & \cos^2 \theta
\end{array}
\right)
\end{equation}
With no loss of generality, we can assume that $\hat{k}=\hat{z}$, so that the limits of integration,
which range only over the illuminated portion of the sphere, are $\theta \in [ \pi/2, \pi]$
and $\phi \in [ 0, 2\pi]$. By inspection, all off-diagonal elements vanish. The diagonal elements
are all equal to $2\pi R^2/3$, so that 
\begin{equation}
{\bf G}=\frac{2}{3} ~\pi R^2 ~{\rm diag} (1,1,1),
\end{equation}
simply the unit tensor times a geometrical factor. Therefore, from Eq.\ \ref{eq:tdepforce}, the force equation is
\begin{equation}
\label{eq:sphmotion}
m \, \frac{d\vec{v}}{dt}=\frac{2}{3}~\pi R^2\, C_m I \, \hat{k}.
\end{equation}
Not surprisingly, the impulse is along $\hat{k}$.
Given the form of this equation, it is fair to say that, in this context,  the effective area
of a sphere is $(2/3) \pi R^2$, i.e., smaller than the geometrical cross-section. 

Since we have assumed that the sphere is a homogeneous solid, we eliminate $R$ in
favor of $m$ and $\rho$, according to
$R=(3m/4 \pi \rho)^{1/3}$, which, after substitution into Eq.\ \ref{eq:sphmotion}, leaves
\begin{equation}
\label{eq:emsph}
\frac{d\vec{v}}{dt}=\biggl( \frac{\pi}{6 \rho^2 m} \biggl)^{1/3} \, C_m I \, \hat{k}.
\end{equation}
which is analogous to the far right-hand side of Eq.\ \ref{eq:cubeforce} for a cube.
This shows that a ``scale-up'' to a larger mass, for a given density, goes as $m^{-1/3}$,
and that $(\rho^2 m)^{-1/3}$ can be seen as an approximation to the {\it effective} area-to-mass ratio,
with a correction term of $(\pi/6)^{1/3}$ for a sphere. Therefore, in considering the relative plausibility
of substantially modifying the orbits of, let us say, a 1 g object and a $10^6$ g object, 
one should work with a scale factor of $10^{-2}$, rather than $10^{-6}$ when 
evaluating the relative velocity change. Moreover, the internal mass distributions of large objects are not homogeneous;
the density $\rho$ appearing in Eq.\ \ref{eq:emsph} is not the material density but rather the average
density of the entire body, thereby adding to the plausibility of maneuvering large objects. 
One caveat here is that, for this kind of scaling, the laser beam must overfill the target,
since the projected area, assumed to be fully illuminated, enters into the calculation.

To touch base with the earlier approximate $\Delta v$ requirement set by Eq. \ref{eq:dvreq}, we
take the two examples thus far derived and consider two homogeneous aluminum targets [$C_m=2$ dynes W$^{-1}$; 
a conservative representative value taken from P.\ Combis, et al.\ (personal communication)],
each with a mass of 100 g. With an average intensity of 100 W cm$^{-2}$ (\S3.1), we find
from Eqs. \ref{eq:cubeforce} and \ref{eq:emsph}, 
\begin{equation}
\Delta v=0.18 \,  \Delta t ~{\rm m ~s}^{-1} ~~{\rm (sphere)} ~~~~~~
\Delta v=0.23 \,  \Delta t ~{\rm m ~s}^{-1} ~~{\rm (cube)} 
\end{equation}
if $\Delta t$ is given in seconds. The difference is a consequence of the fact that
the effective area-to-mass ratio of a cube is larger than that of a sphere of equal mass by the factor $(6/\pi)^{1/3}$.
Thus a 100 km perigee reduction requires an engagement duration of 100-200 seconds for these
two objects, attainable on a single pass, assuming a laser system as outlined in \S2.4, and as discussed
in more detail in \citet{sasha12}. Reacquisition and engagement over multiple orbits may be required
in many cases, where larger perigee changes are desired. 

The previous numerical estimate is provided only as a means to reinforce feasibility studies of the basic methodology.
Obviously, the efficacy of perigee reduction depends on the {\it direction} of $\Delta v$, not just its
magnitude. Suffice to say that for these two shapes $\Delta \vec{v}$ is along the beam, and it has been
shown for this case that the optimum engagement angle is approximately 30$^{\circ}$ uprange of laser zenith \citep{sasha12},
in which case a substantial radial impulse may be imparted to the orbiting debris fragment. In the following examples,
the impulse does {\it not} lie exclusively along the beam and the issue of engagement optimization is an open question.

\subsection{Plate}

The simplest example of an object that will exhibit an off-beam response to laser ablation
is a flat plate oriented such that the surface normal has a component orthogonal to the beam direction.
The situation is as illustrated in Fig.\ 4, with the angle $\phi \in [0,\pi]$ describing the orientation relative
to edge-on. We assume that the plate has, in effect, only two dimensions, i.e., we neglect the edges.
The laser is incident from the left, such that $\hat{k}=\hat{x}$.
The particular shape of the plate used in this example --- circular, square, random --- is not important,
as long as it is thin in one dimension and perfectly flat. Only the surface area matters. 

\begin{figure}
\begin{center}
\includegraphics*[width=8cm,angle=0]{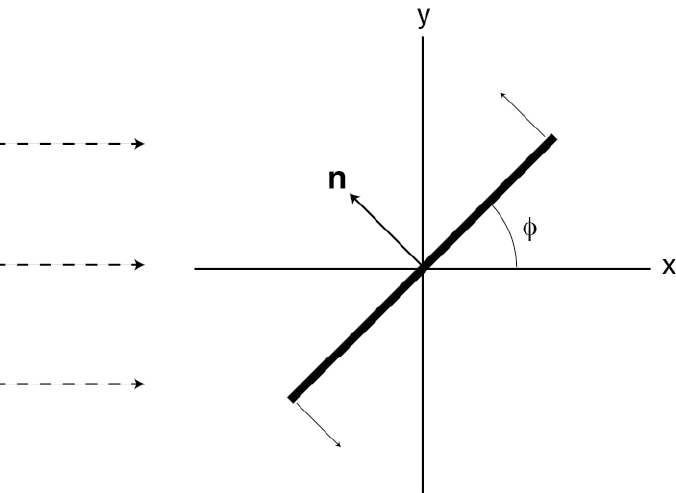}
\end{center}
\caption{Geometric setup showing a thin plate projected onto the $xy$-plane. 
Laser light ({\it dashed arrows}) is incident from the left. As described in \S4.3, we first assume
that the plate is not spinning. In \S4.5, we let the plate spin counter-clockwise at a constant angular velocity
($\omega = \omega \hat{z}$).}
\end{figure}

The surface normal is given by
\begin{equation}
\label{eq:platenormal}
\hat{n}=
\left(
\begin{array}{c}
-\sin \phi\\
\cos \phi
\end{array}
\right),
\end{equation}
where the $z$-component is omitted, since it plays no role in this example.
The area matrix (again, no $z$-components) is, therefore,
\begin{equation}
{\bf G}= A \left(
\begin{array}{cc}
\sin^2 \phi & -\sin \phi \, \cos \phi \\
 -\sin \phi \, \cos \phi & \sin^2 \phi
 \end{array}
 \right),
\end{equation}
from which
\begin{equation}
\label{eq:flatplate}
\left(
\begin{array}{c}
\dot{v}_x \\
\dot{v}_y
\end{array}
\right) =\frac{C_m IA}{m}~
\left(
\begin{array}{c}
\sin^2 \phi \\
-\sin \phi \, \cos \phi
\end{array}
\right),
\end{equation}
or
\begin{equation}
\label{eq:ablateplate}
\frac{d \vec{v}}{dt}=- \frac{C_mIA}{m} ~\sin \phi ~\hat{n}
\end{equation}
Not surprisingly, the trajectories are straight lines with the displacement varying with $t^2$.
The maximum $x$-component of velocity corresponds to the face-on orientation, the maximum
$y$-component of velocity corresponds to $\phi =\pi/4$ or $3\pi/4$, owing to the competing effects of
projected area and alignment of the normal along the $\hat{y}$ direction, and the maximum total
velocity corresponds to the face-on configuration.

\subsection{Plate Responding to Photon Pressure}

At low surface intensities, below the threshold for melting,
the kinematic response of an illuminated object is dominated by photon pressure.
A scheme for space debris maneuvering using photon pressure is presented by \citet{mason}.
Although somewhat off-topic, the situation bears similarities to the pure ablation case,
so we provide a brief treatment here, restricting the discussion to this single example of a flat plate.
The problem is similar to that of pure ablation, although slightly
more complex at the macroscopic level, since the relative amounts of absorption, specular reflection, and diffuse
reflection all play a role; absorption and reflection transmit momentum in two different
directions, unless the beam happens to be aligned with the surface normal.

\begin{figure}
\begin{center}
\includegraphics*[width=12cm,angle=0]{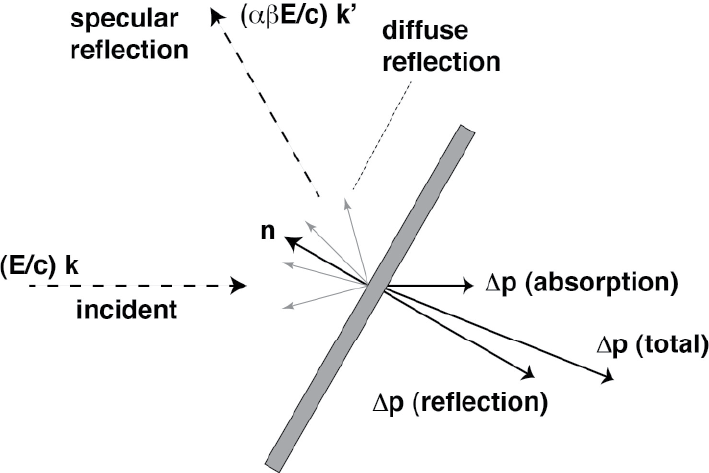}
\end{center}
\caption{Momentum vector relationships for laser illumination of a flat plate in the photon pressure regime.
It assumed that the beam overfills the target. Reflection transfers momentum to a direction opposite
the surface normal ${\bf n}$, while absorption transfers momentum along the incident beam direction ${\bf k}$. 
The magnitude of the momentum transferred along ${\bf k}$ decreases with increasing albedo $\alpha$. 
The quantity $\beta$ is the fraction of reflected light in the specular component.
We assume that the diffuse radiation is hemispherically distributed, with a net vector along the normal.
The figure shows that, typically, the net momentum acquired by the target is neither along the beam
nor precisely opposite the normal.}
\end{figure}

The situation is illustrated in Figure 5. The fraction of incident light that is reflected
is determined by the albedo $\alpha$. The fraction of reflected light that goes into
the specular component is denoted by $\beta$, so that the diffuse reflection fraction is given
by $\alpha (1-\beta)$.
The direction of the specular reflection component is denoted $\hat{k}^{\prime}$ 
and is given in terms of the incident beam direction and the surface normal by
\begin{equation}
\label{eq:kkprime}
\hat{k}^{\prime} = \hat{k}-2 \, (\hat{k} \cdot \hat{n}) \, \hat{n}
\end{equation}
Momentum conservation gives
\begin{equation}
m \,  \frac{d \vec{v}}{dt}= 
\frac{IA}{c} \, \vert \hat{k} \cdot \hat{n} \vert  \biggl[ \hat{k}
- \alpha \beta \, \hat{k}^{\prime}
-\frac{1}{2} \, \alpha \, (1-\beta) ~ \hat{n} \biggr],
\end{equation}
The factor of 1/2 represents our assumption that the diffuse reflection component
originating at a given point on the surface is distributed uniformly into a hemisphere.
Referred to the coordinate system in Fig.\ 4, the $x$ and $y$ force equations are
\begin{equation}
\label{eq:veexdot}
\dot{v}_x = \frac{IA}{mc} \, \sin \phi \, \biggl[ 1-\alpha \beta + \frac{1}{2} \, \alpha (1-\beta) \sin \phi
+ 2 \alpha \beta \sin^2 \phi \biggr]
\end{equation}
\begin{equation}
\label{eq:veeydot}
\dot{v}_y=- \frac{IA}{mc} ~ \alpha \sin \phi \cos \phi ~ \biggl[ \frac{1}{2} \, (1-\beta) + 2\beta \sin \phi \biggr]
\end{equation}
Excepting the case where the beam is aligned with the normal, it is
only in the case of pure absorption ($\alpha=0$) that the momentum transfer is strictly along the beam 
($\dot{v}_y=0$), in which case the force equation becomes
\begin{equation}
\dot{v}_x = \frac{IA}{mc} \, \sin \phi ~~~~({\rm pure ~absorption}).
\end{equation}
At the other extreme --- pure specular reflection ($\alpha=1$, $\beta=1$) --- we find
\begin{equation}
\frac{d\vec{v}}{dt}=2 \, \frac{IA}{mc} ~\sin^2 \phi 
\left(
\begin{array}{c}
\sin \phi \\
-\cos \phi
\end{array}
\right) ~~~~({\rm pure ~specular~reflection}).
\end{equation}
From Eq.\ \ref{eq:platenormal}, we see that
the expression for pure specular reflection can also be written as an acceleration along the normal;
\begin{equation}
\frac{d\vec{v}}{dt} = -2 \, \frac{IA}{mc} \, \sin^2 \phi ~\hat{n}  ~~~~({\rm pure ~specular~reflection}),
\end{equation}
where we recover the magnitude of the recoil ($2IA/mc$) and the
``effective mechanical coupling coefficient'' ($2/c =6.7 \times 10^{-4}$ dyne W$^{-1}$) in the case of normal incidence.
One can compare this expression to the analogous case of pure ablation
(Eq.\ \ref{eq:ablateplate}). The dependence
on the incident angle differs in the two cases. In the photon pressure case,
an additional power of $\sin \phi$ appears owing simply to Snell's Law, whereas
ablation does not obey the rules of optics in this case. Thus the efficiency of the photon pressure
mechanism drops rapidly as grazing incidence ($\phi \rightarrow 0$) is approached.

We note one further effect. Since a fraction of the laser energy is absorbed, the target will re-radiate
this energy. In the rest frame of a continuously irradiated thin plate, with illuminated
surface 1 and the ``dark'' surface 2, 
re-radiation (subscript RR) constitutes an additional contributor to the normal component of the momentum transfer;
\begin{equation}
(\dot{v}_n)_{\rm RR}=-\frac{A}{mc}  ~ (I_1 -I_2).
\end{equation}
Since $I_1 \geq I_2$, this momentum component
is also opposite to the surface normal.

In general, determining $I_1$ and $I_2$, even for this simple geometry,
is a complicated problem in non-linear heat diffusion, and we pursue it no further here.
We can evaluate the two limiting cases: (1) $I_2=I_1$ (no added momentum), and (2) $I_2=0$, in which case
\begin{equation}
\label{eq:rerad}
\biggl( \frac{d\vec{v}}{dt} \biggr)_{\rm RR}= \frac{I_1 A}{mc} ~
\left(
\begin{array}{c}
\sin \phi \\
-\cos \phi
\end{array}
\right).
\end{equation}
The re-radiation intensity is highly material-dependent. In lieu of an exact solution,
the result for the maximum re-radiation impulse (Eq.\ \ref{eq:rerad})
can be added provisionally to Eqs.\ \ref{eq:veexdot} and \ref{eq:veeydot}, where it becomes 
evident that it may be of a magnitude comparable to the other contributors, and favorably modifies 
the photon pressure scheme for engaging space debris. Reliable quantitative evaluations await 
detailed numerical calculations. 

\subsection{Spinning Plate}

Returning now to pure ablation pressure,
suppose the plate described in \S4.3 is spinning around an axis perpendicular to a surface normal, such that the spin axis
intersects the center of mass, as shown in Fig. 4,
and that the spin frequency $\omega$ is unaffected by ablation. Suppressing the $z$-component,
which plays no role here, the surface normal is
\begin{equation}
\hat{n}=
\left(
\begin{array}{c}
-\sin (\omega t+\phi)\\
\cos (\omega t+\phi)
\end{array}
\right)
\end{equation}
The time-dependent area matrix (again, no $z$-components) is, therefore,
\begin{equation}
\label{eq:gmatplate}
{\bf G}(t)= A \left(
\begin{array}{cc}
\sin^2 (\omega t+\phi) & -\sin (\omega t+\phi) \, \cos (\omega t+\phi) \\
 -\sin (\omega t+\phi) \, \cos (\omega t+\phi) & \sin^2 (\omega t+\phi) 
 \end{array}
 \right)
\end{equation}
from which
\begin{equation}
\left(
\begin{array}{c}
\dot{v}_x \\
\dot{v}_y
\end{array}
\right) =\frac{C_m IA}{m}~
\left(
\begin{array}{c}
\sin^2 (\omega t+\phi)\\
-\sin (\omega t+\phi) \, \cos (\omega t+\phi)
\end{array}
\right)
\end{equation}
With the initial condition $(v_x,v_y)=(0,0)$, the $x$ and $y$ velocity components are
\begin{equation}
\label{eq:vxspin}
v_x=\frac{C_mIA}{2m \omega} ~[\, \omega t - \sin \omega t \, \cos (\omega t + 2 \phi) \, ]
\end{equation}
\begin{equation}
\label{eq:vyspin}
v_y=- \frac{C_mIA}{2m \omega} ~\sin \omega t \, \sin (\omega t + 2 \phi)
\end{equation}
By spin-averaging the previous expression for $v_y$, we find that
\begin{equation}
\label{eq:vyavg}
\langle v_y \rangle=- \frac{C_mIA}{4m \omega} ~ \cos 2 \phi,
\end{equation}
which can be thought of as an average drift velocity --- the presence of spin does not generally obviate
the need for considering off-beam velocity components.
Although the $y$-component of the impulse
is indeed oscillatory, with an average magnitude of zero, the target drifts.
However, the resultant trajectory is sensitive to the value of $\phi$ at the onset of ablation.
When $\phi=(2n-1) \, \pi/4$ ($n=1,2,3,...$), the drift term vanishes. 
In all other cases, either a positive or negative drift
velocity orthogonal to the beam is present.

\begin{figure}
\begin{center}
\includegraphics*[width=10cm,angle=0]{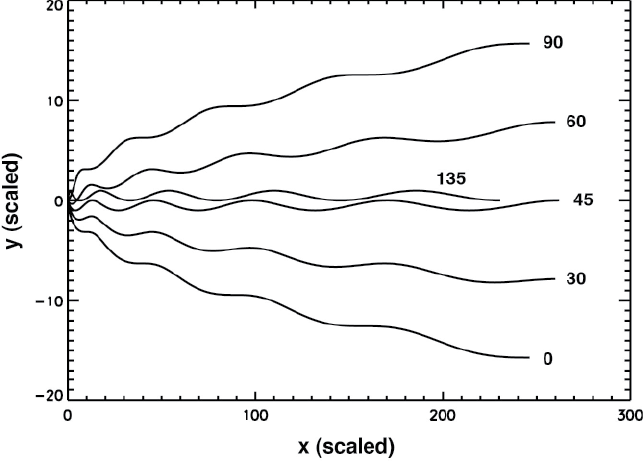}
\end{center}
\caption{Trajectories of the center of mass of a spinning plate for six different initial 
phase angles (as indicated), covering 2.5 rotation cycles. Axes are scaled (see text).}
\end{figure}

The trajectories are approximately parabolic $(x \propto y^2$), as can be inferred
from Eqs.\ \ref{eq:vxspin} and \ref{eq:vyavg}, i.e., if ablation continues
indefinitely, then the trajectory approaches a parabola, with linear eccentricity $q$ (defined
so that $y^2 = 4qx$) given by
\begin{equation}
q=\frac{C_mIA \, \cos^2 2\phi}{16 m \omega^2}
\end{equation}
which shows that the parabola ``narrows'' as $\omega^{-2}$ --- for sufficiently large $\omega$,
a straight-line approximation becomes progressively more tenable. On shorter time scales, the
trajectories are somewhat complex, and depend sensitively on the phase angle, as shown in Fig.\ 6.
Note also the lack of trajectory symmetry; comparing $\phi=45^{\circ}$ and $\phi=135^{\circ}$,
the former travels farther, since it gets a ``head start'' at early times, rotating into a more complete alignment
with the laser field.

\subsection{Cylinder}

We now examine a solid, homogeneous cylinder of mass density $\rho$, with
radius $r$ and height $h$, supposing that the cylinder has circular end caps
of the same material as the cylinder walls. 
A  particular mode of cylinder rotation will be considered, but first, to calculate ${\bf G}$, 
let the cylindrical axis be aligned with the $z$-axis. After determining ${\bf G}$
in this orientation, the more general form is found through a rotational transformation below.
Using the standard cylindrical
coordinates $(\varrho, \phi, z)$, we have
$\hat{\varrho}=(\cos \phi, \sin \phi, 0)$. The area matrix can be constructed
as the sum of two components, with only one (or neither) end cap illuminated.

\begin{equation}
{\bf G}=\pi r^2 \,  \hat{z} \hat{z} +\int dA ~\hat{\varrho}\hat{\varrho}
\end{equation}
Since
\begin{equation}
\hat{\varrho}\hat{\varrho}=
\left(
\begin{array}{ccc}
\cos^2 \phi            & \sin \phi \cos \phi & 0 \\
\sin \phi \cos \phi & \sin^2 \phi             & 0 \\
0                             & 0                             & 0
\end{array}
\right),
\end{equation}
and we integrate $z$ over $[-h/2, h/2]$ and $\phi$ over $[\pi, 2\pi]$, we find
\begin{equation}
{\bf G}=\frac{\pi}{2}
\left(
\begin{array}{ccc}
rh    & 0        & 0     \\
0     &     rh   & 0     \\
0     & 0         & 2r^2
\end{array}
\right)
\equiv \pi r^2
\left(
\begin{array}{ccc}
a   & 0  & 0 \\
0   & a  & 0 \\
0   & 0  & 1
\end{array}
\right),
\end{equation}
where $a \equiv h/2r$. Note that the special case $h=2r$ ($a=1$) leaves ${\bf G}$
as a multiple of the unit tensor, so that the recoil is strictly along the beam in this special case, 
regardless of orientation, i.e., regardless of the $\hat{k}$-direction, the recoil will always
be along $\hat{k}$. In fact, the two ranges $a<1$ and $a>1$ result in qualitatively different
responses, which motivates the introduction of $a$. 

\begin{figure}
\begin{center}
\includegraphics*[width=10cm,angle=0]{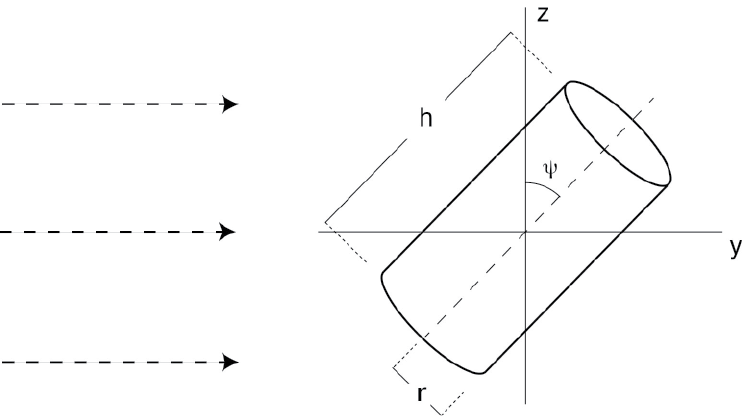}
\end{center}
\caption{Relative orientation of a solid cylinder with respect to a laser beam
propagating along the $y$-axis,
described by the angle $\psi$, with the cylindrical axis restricted to the $yz$-plane. 
At most, one end cap is illuminated. With the orientation
and dimensions shown, the recoil will have a positive $y$-component and a negative
$z$-component. As the $h/r$ ratio is reduced the $z$-component switches from negative to
positive, passing through zero when $h=2r$ ($a=1$).}
\end{figure}

Let the laser propagate along the $y$-axis ($\hat{k}= \hat{y}$). Rather than allowing
freedom in $\hat{k}$,
we introduce a degree of freedom to allow for a range of cylinder orientations, viz., 
we allow rotation about the $x$-axis by an angle $\psi \in [0,\pi]$ (see Fig. 7),
which may be time dependent. In practice, it may be easier to calculate the recoil
by fixing the object and ``moving'' the laser. However, when coupled with the orbital
calculations, it is more natural to fix the laser in an Earth-centered coordinate system
and let the object rotate in this system. Although the
orbital mechanics aspects of the problem are being deferred to a later paper, we choose that approach here.

To rotate the corresponding area matrix in this {\it fixed} coordinate system, such that
the cylindrical axis always lies in the $yz$-plane, we use
the transformation ${\bf G}_{\rm rot}=R{\bf G}R^{-1}$, where the rotation matrix $R$ is
\begin{equation}
R=
\left(
\begin{array}{ccc}
1     & 0                 & 0     \\
0     &   \cos \psi   &  \sin \psi     \\
0     &   -\sin \psi   & \cos \psi
\end{array}
\right).
\end{equation}
The rotated area matrix is thus given by
\begin{equation}
{\bf G}_{\rm rot}=\pi r^2
\left(
\begin{array}{ccc}
a     & 0                                                 & 0     \\
0     &     a \, \cos^2 \psi + \sin^2 \psi   & -(a-1) \sin \psi \cos \psi     \\
0     &   -(a-1) \sin \psi \cos \psi         & a \, \sin^2 \psi + \cos^2 \psi
\end{array}
\right).
\end{equation}
Using our definition of $a$, we eliminate $r=(m/2\pi \rho a)^{1/3}$, so as to recast the following 
development in terms of $(\rho^2 m)^{-1/3}$, as above, and the equation of motion is
\begin{equation}
\label{eq:decyl}
\frac{d\vec{v}}{dt}=\biggl( \frac{\pi}{4 \rho^2 ma^2} \biggr)^{1/3} ~C_mI
\left(
\begin{array}{c}
0 \\
a \, \cos^2 \psi + \sin^2 \psi \\
-(a-1) \sin \psi \cos \psi
\end{array}
\right).
\end{equation}

The $y$-component of force, hence velocity, is always positive, whereas the $z$-component may
be positive, negative, or zero.
There are two special cases: $\psi=0$ and $\psi=\pi/2$. If $\psi=0$, the endcaps are not illuminated,
and there is no $z$-component to the impulse. For the $y$-component, the impulse can, by analogy
to the spherical case discussed earlier, be associated with an effective cross-section, which in this case
is $(\pi/2) \, rh$, again smaller than the projected area $2rh$. When $\psi=\pi/2$, only the endcap is illuminated,
and again there is no $z$-component to the impulse vector.
As for any flat surface whose normal is anti-parallel to $\hat{k}$, the effective area is equal to its
actual geometrical area of $\pi r^2$. Other than for the ``sphere-like''
$a=1$ case, if $\psi$ is not an integer multiple of $\pi/2$, the force equation shows a $z$-component, 
and the motion is not parallel to $\hat{k}$.

Proceeding, let us consider a case in which the cylinder rotates about the $x$-axis at a constant
angular velocity $\omega$ (clockwise in Fig.\ 7), so that $\psi =\omega t + \phi$, where $\phi$ sets the
orientation at the moment the laser first illuminates the cylinder; for reference, if $\phi =0$ at $t=0$, the initial
orientation is such that the cylindrical axis is aligned with the $z$-axis. 

With the initial conditions $v_y =v_z =0$, the velocity solutions of Eq.\ \ref{eq:decyl} are
\begin{equation}
\label{eq:vyosc}
v_y=2 ~ \frac{a+1}{a^{2/3}} \,  \kappa \omega \,  \biggl[ \omega t +\frac{a-1}{a+1}
\, \sin \omega t \, \cos (\omega t +2 \phi) \biggr]
\end{equation}
\begin{equation}
\label{eq:vzcyl}
v_z= -2 ~ \frac{a-1}{a^{2/3}} \, \kappa \omega \, \sin \omega t \, \sin (\omega t + 2 \phi)
\end{equation}
where $\kappa$, with the dimension of length, is defined by
\begin{equation}
\kappa \equiv \frac{C_mI}{4\omega^2} \, \biggl( \frac{\pi}{4\rho^2 m} \biggr)^{1/3}.
\end{equation}
In a manner similar to the case of a rotating plate, the rotating cylinder has a drift
velocity in a direction orthogonal to the beam: from Eq. \ref{eq:vzcyl}, it is easy to show that
the spin-averaged drift velocity is
\begin{equation}
\langle v_z \rangle =-\frac{a-1}{a^{2/3}} ~\kappa \omega \cos 2 \phi,
\end{equation}
which vanishes only when $\phi=(2n-1) \pi/4$ ($n$=1, 2, 3,...), identical to the case of a plate.
The magnitude of the drift velocity is at its maximum whenever $\phi$ is an integer multiple
of $\pi/2$. The leading term for $v_y$ (along the beam) is independent of $\omega$, while
$\langle v_z \rangle \propto \omega^{-1}$.

Neglecting the oscillatory second term of Eq. \ref{eq:vyosc}, one finds that the minimum
$y$-component of the velocity occurs for the case $a=2$, i.e., for a cylinder with the approximate relative dimensions
found for a (U.S.A. standard) soda can. This detail is related neither to the $a$-dependence
of the average projected area, which decreases monotonically as $a^{-2/3}$ for a fixed volume,
nor to the minimum absolute area for a fixed volume ($a$=1). The {\it maximum} attainable velocity
corresponds to either limit $a \rightarrow 0$ or $a \rightarrow \infty$, which becomes unphysical, i.e.,
an infinite disk of vanishing thickness in the former case, or an infinitely long rod of vanishing radius in the latter.

\begin{figure}
\begin{center}
\includegraphics*[width=10cm,angle=0]{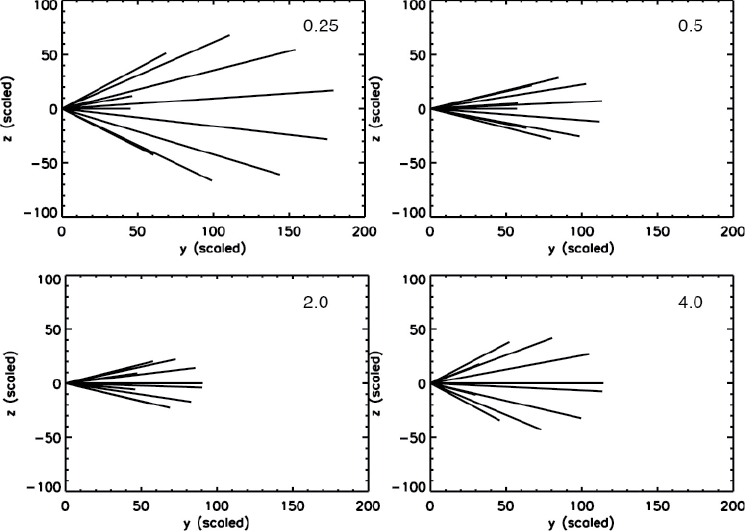}
\end{center}
\caption{Trajectories of non-spinning, solid cylinders for four values of $a=h/2r$ (as indicated in upper right
of each panel), for equal-mass cylinders of height 
$h$ and radius $r$, for ten different orientations. Axes are scaled, as described in text.}
\end{figure}

With the initial conditions $y=z=0$ for the cylinder's center of mass, the coordinate
solutions of Eq.\ \ref{eq:decyl} are
\begin{equation}
y(t)= \frac{a+1}{a^{2/3}} \, \kappa  \, \biggl[ (\omega t)^2 -  \frac{a-1}{a+1}
\biggl(\omega t \, \sin 2\phi - \sin \omega t \, \sin (\omega t + 2\phi )  \biggr) \biggr]
\end{equation}
\begin{equation}
z(t)=- \frac{a-1}{a^{2/3}} \, \kappa \, 
[ \omega t \,  \cos 2\phi - \sin \omega t \,  \cos(\omega t + 2\phi)]
\end{equation}
As expected, spin has no effect on the center-of-mass motion when $a=1$, since then
all terms containing $\omega$ drop out. In the limit $\omega t \gg 1$ the trajectories
become parabolic.

The behavior of a non-spinning cylinder ($\psi$ fixed) is found by taking the limit 
$\omega \rightarrow 0$ (also, $\psi = \phi$ in this case),
\begin{equation}
\label{eq:vycyl}
v_y(t; \omega=0)= \biggl( \frac{ \pi}{4  \rho^2 m}\biggr)^{1/3}
 ~ \biggl [\frac{a+1+(a-1) \, \cos 2\psi}{2a^{2/3}} \biggr ]  ~C_mIt
\end{equation}
\begin{equation}
v_z(t; \omega=0)=- \biggl( \frac{ \pi}{4 \rho^2 m}\biggr)^{1/3} ~
\frac{(a-1) \sin 2\psi}{2a^{2/3}} ~C_mIt
\end{equation}
With the same initial conditions $y=z=0$, the coordinate solutions are
\begin{equation}
y(t; \omega=0)=\biggl(\frac{\pi}{32 \rho^2 m} \biggr)^{1/3} ~ \frac{ a+1+(a-1) \, \cos 2\psi}{2a^{2/3}} ~C_m It^2
\end{equation}
\begin{equation}
z(t; \omega=0)=-\biggl(\frac{\pi}{32 \rho^2 m} \biggr)^{1/3} ~ \frac{(a-1) \sin 2\psi}{2a^{2/3}}  ~C_m I t^2
\end{equation}
(Of course, the non-spinning solution may also be found directly by solving
Eq.\ \ref{eq:decyl} with $\psi$ fixed.) The trajectories are straight lines for all $t$, but have
a marked dependence on both $a$ and $\psi$, as shown in Fig.\ 8..

Recalling that the $a=1$ case leads to ``sphere-like'' motion of the cylinder, it should be made clear
that this is true only
in the sense that the recoil is along $\hat{k}$ in either case. Comparing the recoil velocities, we see
from Eqs.\ \ref{eq:emsph} and \ref{eq:vycyl} that the velocity of the $a=1$ cylinder is $(3/2)^{1/3}
\approx 1.14$ times that of a sphere, assuming identical masses, mass densities, and coupling coefficients.

\section{Transfer of Spin Angular Momentum}

In the previous section, we added spin to a thin plate and to a cylinder, assuming that the spin is
independent of the laser/target interaction. In this section, we examine the interaction
directly, with the aim of characterizing the extent to which laser ablation induces 
or otherwise affects spin. While it is unlikely that {\it all} potential targets are spinning ---
for example, spin may be damped by Earth's magnetic field \citep{praly} --- some are.
Moreover, in some cases, the laser interaction may itself induce spin.

Since ablation causes a local force per unit area  on an object, a torque about the
center of mass may result. Here we assume that the force points exactly opposite the local
surface normal, as in earlier sections. If the areal summation of these local torques
is non-zero, the spin angular momentum of the object may be altered. We refer to this as the ``static torque'' 
(symbolized by $N_s$), since it applies whether or not the target is spinning at the onset of the laser interaction. 
The symmetry aspects of the target figure
prominently here, since certain symmetries ensure that for every torque there is 
another that is equal but opposite, and spin cannot be induced; for example, a circular
disk, or a uniform rod. However, with the exception of NaK spheres, there is no reason 
to expect that space debris possess {\it any} kind of symmetry.

The static torque is given by 
\begin{equation}
\label{eq:statictorque} 
{\vec N}_s=C_m I \, \int dA ~(\hat{k} \cdot \hat{n}) ~ ( {\vec r} \times  \hat{n} ) 
\end{equation}
where $\vec{r}$ is the position vector of a surface element relative to a convenient
origin, usually the center of mass. Note that we cannot use the area matrix here, since
the cross-product must be taken inside the integral.
From Eq.\  \ref{eq:statictorque} we can make an order-of-magnitude estimate of the typical
torque on an irregularly-shaped object, providing that all symmetry effects are disregarded.
Approximating the moment of inertia by $m R^2$, where $R$ is a characteristic dimension
of the object, we can write $\dot{\omega} \sim C_mIR/m$. 
Suppose $C_m=1$ dyne W$^{-1}$, $I=100$ W cm$^{-2}$ (again, a time-averaged intensity), $R=10$ cm, and
$m=100$ g. Then, after one second, the induced spin frequency is approximately 10 rad s$^{-1}$.
Of course, we cannot simply ignore torque-canceling symmetries, but this
clearly shows that spin effects could play a role in laser/target interactions.
A few examples are provided in the remainder of this section.

A second source of torque may arise if an object is already spinning, or if it is set to
spinning as a result of asymmetry, as just discussed. In this case the ablation
flow, viewed from the observer (inertial) frame is not purely normal with respect to a given
surface, but contains both a normal component and an orthogonal $\vec{\omega} \times \vec{r}$ component.
We refer to the  torque imparted to the object from the effect of spin as the ``kinematic torque,'' and
symbolize it by $N_k$.
To get an order-of-magnitude estimate of this effect, we take the azimuthal velocity component of the ablation
flow to be $\sim \omega R$. The rate of angular momentum creation in the flow is therefore
approximately $\omega R^2$ times the mass loss rate, which is $\sim C_mI R^2/v_a$ (with $v_a$ the
ablation flow speed in the frame of the ablating surface), thus
producing a back-reaction torque on the spinning object $N \sim C_m I R^4 \,  \omega / v_a$,
with the result that $\dot{\omega} \sim -(C_m I R^2
/mv_a) \, \omega$, thus showing an exponential spin damping over the time scale $mv_a/ C_m I R^2$.
A more careful estimate would include the change to the moment of inertia from mass loss as ablation proceeds.
In fact, as we show below, the mass loss effect on the moment
of inertia may exactly compensate the torque, leaving the spin unaffected.

Formally, the kinematic torque can be calculated according to
\begin{equation}
\label{eq:spintorque}
 {\vec N}_k=\int dA ~\frac{d \dot{m}_a}{dA} ~{\vec r} \times (\vec{\omega} \times {\vec r})
\end{equation}
where we make the assumption that the local areal mass loss rate is given by
\begin{equation}
\label{eq:mdotgeneral}
\frac{d \dot{m}_a}{dA}= \frac{C_mI \, \hat{k} \cdot \hat{n}}{v_a}.
\end{equation}
which is simply an expression of linear momentum conservation under the conditions that
the ablation flow is parallel to the normal vector of the differential surface element in its rest frame, and that
the momentum impulse is precisely anti-parallel to the surface normal. It can also be taken as a
definition of the mechanical coupling coefficient $C_m$.

Given an inertia tensor ${\bf J}$, the torque equation (angular momentum transfer) is
\begin{equation}
\label{eq:spindyn}
\frac{d}{dt} \, ({\bf J} \cdot  \vec{\omega})= \vec{N},
\end{equation}
where the total torque is given by
\begin{equation}
\label{eq:totaltorque}
{\vec N}={\vec N}_s + {\vec N}_k =C_m I \, \int dA ~\hat{k} \cdot \hat{n} ~ \biggl[ {\vec r} \times  \, 
\biggl( \hat{n}  + \frac{\vec{\omega} \times \vec{r}}{v_a} \biggr) \biggr],
\end{equation}
and we remind the reader that the integral is to be taken only over the illuminated portion of the target.

We work out a few idealized examples below. Ultimately, our goal
is to determine the manner in which target spin might affect engagement strategies, if at all. Here, however,
we focus on providing a few examples in the interest of imparting some sense of the scope of the problem. 

\subsection{Mass Loss}

As discussed earlier, we follow the standard practice of ignoring
mass-loss effects on the inertia of the debris target. We adopt the same procedure
for dealing with spin changes induced by a static torque. However, in treating kinematic
torque, we are required to treat the mass as a time-dependent quantity. These latter two
choices are justified in this subsection.

For the static torque, a scalar approximation to Eq.\ \ref{eq:statictorque} is
\begin{equation}
\label{eq:scalarstatic}
\frac{d}{dt} \, (J\omega)_S  \sim \dot{m}_a v_a R,
\end{equation}
which bears a close resemblance to the linear momentum transfer equation
\begin{equation}
\frac{d}{dt} \, (mv)  \sim \dot{m}_a v_a.
\end{equation}
Equation \ref{eq:scalarstatic} can be thought of as leading to a ``rotational rocket equation''
when mass loss is treated explicitly. However, in the same way that mass is treated as a constant
when working with linear momentum transfer through ablation, we assume a time-constant moment
of inertia when working with rotation under the influence of a static torque, and obtain results to the same order
of accuracy, i.e., the first-order correction term is $O(\Delta m/m_o)$.

A similar scalar representation of the kinematic torque equation, from Eq.\ \ref{eq:spintorque}, is
\begin{equation}
\frac{d}{dt} \, (J\omega)_K  \sim \dot{m}_a \omega R^2,
\end{equation}
which expands to
\begin{equation}
\dot{\omega}  \sim \frac{\dot{m}_a R^2 - \dot{J}}{J} ~ \omega.
\end{equation}
The fraction on the right hand side of this equation is of the same order as $\dot{m}_a/m_o$, so that
$\Delta \omega/\omega_o \propto \Delta m/m_o$ --- rotation under the predominant influence of a kinematic torque
demands an explicit accounting of the mass loss.

The relative magnitudes of the kinematic term and the static term are given by the ratio $ \omega R/v_a$. Since this
ratio is typically small compared to unity, the kinematic torque can, to first order, be ignored when
a static torque is present.  
However, objects possessing a high degree of symmetry are often immune
to static torques, whereas the kinematic torque is comparatively ubiquitous. In such cases it may be of
interest to calculate the rotational dynamical effects of kinematic torque. We work through two examples below.

To summarize, working to leading order in $\Delta m/m$, we can treat the debris mass as a constant 
for the following two cases: (1) linear momentum transfer and (2) changes to rotation under the influence 
of a static torque. On the other hand, the mass should be treated in a time-dependent manner when
evaluating rotational changes under the sole influence of a kinematic torque.

\subsection{Kinematic Torque on a Spinning Cylinder}

An example of kinematic torque is provided by a cylinder of radius $r_c$ and height $h$, spinning on its main
axis, and illuminated only on its curved surface, as illustrated in Fig.\  9. From Eq.\ \ref{eq:statictorque}
it is easy to show that the static torque term vanishes. 

We need the following vectors:
\begin{equation}
\hat{k}=\hat{y} ~~~~ \vec{\omega}=\omega \, \hat{z} ~~~~
\hat{n}=\left(
\begin{array}{c}
\cos \phi \\
\sin \phi \\
0
\end{array}
\right) ~~~~
\vec{r}=\left(
\begin{array}{c}
r_c \cos \phi \\
r_c \sin \phi \\
z
\end{array}
\right)
\end{equation}
where $\phi$ is the standard azimuthal angle measuring angular displacement with respect to
the positive $x$-axis.
Given the initial condition, only the $z$-component is relevant. Therefore, we will need to solve
\begin{equation}
\label{eq:scalartorque}
\dot{J}_{zz} \,  \omega +J_{zz} \, \dot{\omega}=N_z,
\end{equation}
where $J_{zz} \equiv J=(1/2)Mr_c^2$. Eliminating $r_c$ according to $r_c=(M/\pi \rho h)^{1/2}$, 
we have $J=M^2/2 \pi \rho h$ and $\dot{J}=M\dot{M}/\pi \rho h$, which allows us to write
Eq.\ \ref{eq:scalartorque} as
\begin{equation}
\label{eq:scalartorque2}
\frac{M \dot{M}}{\pi \rho h} \, \omega + \frac{M^2}{2\pi \rho h} \, \dot{\omega} = N_z
\end{equation}
We let both $M$ and $\dot{M}$ be functions of time, accounting explicitly for the reduction
of mass and the shrinking target cross-section as ablation proceeds.

\begin{figure}
\begin{center}
\includegraphics*[width=10cm,angle=0]{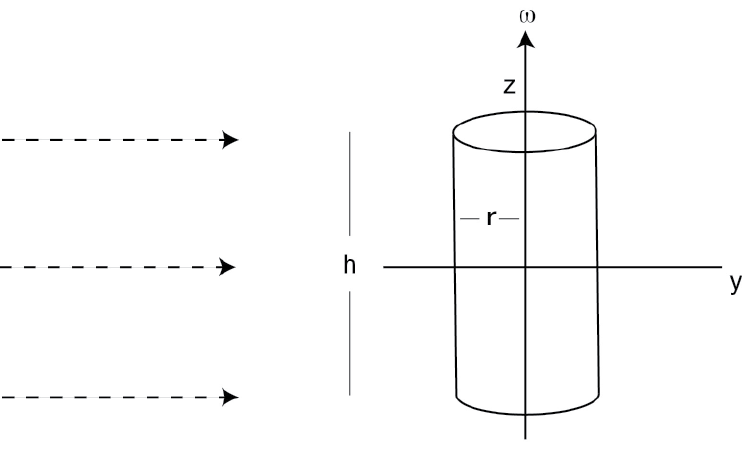}
\end{center}
\caption{
Spinning cylinder of height $h$ and radius $r$, illuminated from the left  ({\it dashed arrows}),
with the laser propagation vector ($\hat{k}=\hat{y}$) perpendicular to the spin axis ($\vec{\omega}=\omega \hat{z}$). }
\end{figure}

The mass loss rate is found from Eq.\ \ref{eq:mdotgeneral}, which takes the form
\begin{equation}
\label{eq:mdotcylinder}
\dot{M}=\frac{C_m I hr_c(t)}{v_a} ~\int_{\pi}^{2 \pi} d\phi ~\sin \phi
=-\frac{2C_m I hr_c(t)}{v_a}.
\end{equation}
We allow for mass loss to reduce only the radius of the cylinder, hence $r_c \rightarrow r_c(t)$,
but assume that $h$ is a constant, i.e., no endcap illumination for this example.

By again re-expressing $r_c$ in terms of $M$, Eq.\ \ref{eq:mdotcylinder} becomes
\begin{equation}
\dot{M}=-\frac{2C_m I h}{v_a} \, \biggl(\frac{M}{\pi \rho h} \biggr)^{1/2},
\end{equation}
from which
\begin{equation}
\label{eq:mmdotcyl}
M= M_o \, (1-\beta t)^2 ~~~~ \dot{M}=-2  M_o \beta  \, (1-\beta t)
\end{equation}
for an initial mass $M_o$, and
where we define $\beta$, with dimension of inverse time, according to
\begin{equation}
\beta \equiv \frac{C_mI}{v_a} \, \biggl( \frac{h}{\pi \rho M_o} \biggr)^{1/2}.
\end{equation}

Next, to find the torque acting on the cylinder, we use Eq.\ \ref{eq:totaltorque}, which, recalling
that the static torque is zero, becomes
\begin{equation}
\vec{N}= \frac{C_mI}{v_a} ~\int dA ~\hat{k} \cdot \hat{n}~ [r^2 \vec{\omega} -
(\vec{\omega} \cdot \vec{r}) \, \vec{r} \, ],
\end{equation}
(using the BACCAB rule for the triple vector product, and also noting that $r^2=r_c^2 + z^2$) with the result
\begin{equation}
\label{eq:ztorque}
N_z= - \frac{2M_o^2 \beta}{\pi \rho h}~(1-\beta t)^3 \, \omega.
\end{equation} 
The torque, as expected, depletes the cylinder of angular momentum, while the magnitude
of the torque decreases with time, since the target cross-section is shrinking.

Finally, to find the net effect on the spin, we combine Eqs.\ \ref{eq:scalartorque2}, \ref{eq:mmdotcyl}, and \ref{eq:ztorque}
to give
\begin{equation}
\label{eq:ddtiomega}
- \frac{2M_o^2 \beta}{\pi \rho h} ~(1-\beta t)^3 \, \omega
+\frac{M_o^2}{2 \pi \rho h}~(1-\beta t)^4 \, \dot{\omega}
=- \frac{2M_o^2 \beta}{\pi \rho h}~(1-\beta t)^3 \, \omega.
\end{equation}
Since the first term on the left-hand-side of Eq.\ \ref{eq:ddtiomega} (reduction of the moment of inertia
through mass loss) entirely accommodates
the torque (right-hand side), we find $\dot{\omega}=0$, i.e., the spin is unaffected by laser irradiation in this configuration. 
In short, the cylinder's loss of angular momentum through mass loss is sufficient to balance
the torque acting on the cylinder. To take another example, although somewhat more involved, it is straightforward
to show that the same effect holds for a right circular cone spinning about its main axis, if illuminated down the axis.

\subsection{Kinematic Torque on a Spinning Sphere}

We showed in the previous example that laser irradiation does not add or subtract spin to a side-illuminated
spinning cylinder. In this subsection, we replace the cylinder with a uniform sphere of radius $R$, with
the spin axis again aligned with the $z$-axis, and the laser propagating along the positive $y$-axis (equatorial illumination),
then perform the same calculation (see Fig.\ 9 for reference).

The mass and mass loss rate as functions of time are derived in a manner similar
to the previous section, i.e., using $J=(2/5) \, MR^2$ and $R=(3M/4\pi \rho)^{1/3}$, with the results
\begin{equation}
\label{eq:mmdots}
M=M_o \, (1-\beta_s t)^3 ~~~~~ \dot{M}=-3M_o \beta_s \, (1- \beta_s t)^2
\end{equation}
where $\beta_s$ is analogous to the decay constant $\beta$ used for the cylinder in the
previous section;
\begin{equation}
\beta_s=\frac{C_mI}{2v_a} \, \biggl( \frac{\pi}{6 \rho^2 M_o} \biggr)^{1/3}.
\end{equation}
Then, using the expressions for $M$ and $\dot{M}$ from Eq.\ \ref{eq:mmdots},
\begin{equation}
\label{eq:iomegadotsphere}
\frac{d}{dt} \, (J\omega) =-2 \, \biggl(\frac{3}{4\pi \rho} \biggr)^{2/3} \, M_o^{5/3} (1-\beta_s t)^4 \biggl[ \beta_s \,  \, \omega
- \frac{1}{5} \,  (1-\beta_s t) \, \dot{\omega} \biggr].
\end{equation}
From Eq.\ \ref{eq:totaltorque}, again discarding the static term, the torque on the sphere is
\begin{equation}
\label{eq:torquesphere}
N_z =-\frac{9}{4} \, \biggl(\frac{3}{4\pi \rho} \biggr)^{2/3} \, M_o^{5/3} \beta_s \, (1-\beta_s t)^4 \, \omega
\end{equation}
Equating the right-hand side of Eq.\ \ref{eq:iomegadotsphere} to the right-hand side Eq.\ \ref{eq:torquesphere},
canceling common factors, gives
\begin{equation}
-2 \beta_s \, \omega +\frac{2}{5} \, (1-\beta_s t ) \, \dot{\omega}= -\frac{9}{4} \, \beta_s \, \omega,
\end{equation}
whose solution is
\begin{equation}
\omega= \omega_o \, (1-\beta_s \, t)^{5/8}.
\end{equation}
Thus the laser will cause the sphere to spin down. The mathematical solution is only an approximation, of course, since
the target will not maintain a perfectly spherical shape, given that the equator loses mass more readily
than the polar regions in this orientation.

\subsection{Asymmetric Dumbbell: An Ablation-Driven Pendulum}

Referring to Fig.\ 10, consider two spherical masses, with radii $R_1$ and
$R_2$ ($R_1 > R_2$), identical mass density $\rho$ (so that $M_1 > M_2$), 
positioned at $\vec{r}_1$ and  $\vec{r}_2$, connected by a massless rod,
with the separation of the centers of the spheres $r_1+r_2 \equiv a$, 
and irradiated such that $\hat{k}=-\hat{x}$. To keep things simple, 
we make the assumption that each sphere has a radius
substantially smaller than the separation distance of the two spheres, so that the 
scalar moment of inertia is simply $ J=\mu a^2$ for the reduced mass $\mu$. 
We ignore the change to the moment of inertia caused by mass loss as per the discussion in \S5.1. 
Also, we ignore the shadowing of one sphere by the other.

\begin{figure}
\begin{center}
\includegraphics*[width=10cm,angle=0]{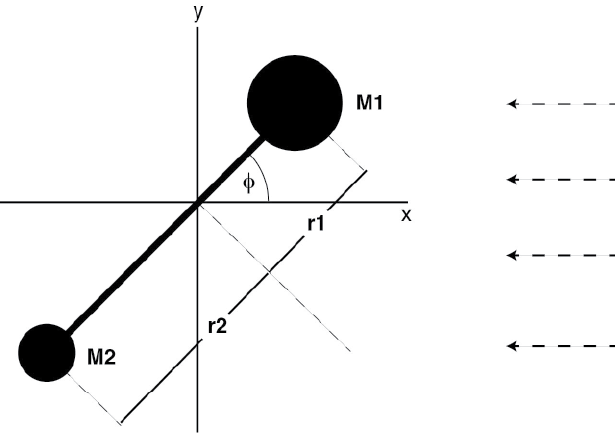}
\end{center}
\caption{Geometric relations for an asymmetric dumbbell consisting of two spheres of identical material
but different radii ($R_1 > R_2$, hence $M_1 > M_2$). The rotation vector lies along the $z$-axis, with rotation about
the system center of mass, which coincides with the origin. The angular position is described
by the angle $\phi$, as shown. Laser is incident from the right ({\it dashed arrows}). }
\end{figure}

With the approximation that the static torque applies only to the spheres, and not to the rod,
we have
\begin{equation}
{\vec N}_s=C_m I \, \sum_{\alpha} \, \int dA_{\alpha} ~(\hat{k} \cdot \hat{n}_{\alpha}) ~ 
( {\vec r}_{\alpha} \times  \hat{n}_{\alpha} ), 
\end{equation}
where $\alpha$ subscripts the two spheres 1 and 2.
Relying on our approximation that $R_{\alpha}$ is small compared to $r_{\alpha}$,
this can be approximated by
\begin{equation}
{\vec N}_s=C_m I \,   \sum_{\alpha} \,  {\vec r}_{\alpha} \times
(\hat{k} \cdot  {\bf G}_{\alpha} ) =
\frac{2}{3} \, \pi C_m I \, \sum_{\alpha} \,  R^2_{\alpha} \, {\vec r}_{\alpha} \times \hat{k},
\end{equation}
where ${\bf G}_{\alpha}=(2\pi/3) R^2_{\alpha} \, {\bf I}$, using our result from \S4.2.
The torque has only a $z$-component, which is, after evaluating the previous expression,
\begin{equation}
N_z= - \frac{2}{3} \, \pi C_m I \, (r_2 R_2^2 - r_1 R_1^2) \, \sin \phi.
\end{equation}
Using $R_{\alpha}=(3m_{\alpha}/4 \pi \rho)^{1/3}$, $r_1=(m_2/m) \, a$,
and  $r_2=(m_1/m) \, a$, the equation of motion becomes
\begin{equation}
\label{eq:pendulum}
\ddot{\phi} + \Omega^2 \, \sin \phi =0,
\end{equation}
where 
\begin{equation}
\Omega^2 =\biggl(\frac{\pi}{6 \rho^2} \biggr)^{1/3}  \, \frac{C_m I}{a} ~
\biggl( \frac{1}{m_2^{1/3}} - \frac{1}{m_1^{1/3}} \biggr),
 \end{equation}
 which is mathematically identical to the non-linear pendulum equation. 
 
 While the governing equation of motion is the same as for a pendulum in a uniform
 gravitational field, the analogy is not perfect. For one thing, if our two spherical
 masses were connected by a rod of negligible mass, and allowed to rotate
 about the center of mass in a gravitational field, there would be no rotation other
 than that imparted by some other means, since the gravitational torques would precisely cancel
 each other. Also, the ``natural frequency'' $\Omega$ is not an intrinsic property
 of the dumbbell configuration --- it is not ``natural'' at all --- since it scales with 
 $C_m^{1/2}$, and has a $\rho$-dependence, as well, whereas the motion of a 
 pendulum is not related to its composition in any way. However, we note that the linear scale
 $a$ plays the same role as the length of a pendulum arm, and, crudely speaking,
 the quantity $C_mI/(\rho^2 m)^{1/3}$ is analogous (and dimensionally equivalent) to the gravitational acceleration.
 
 Equation \ref{eq:pendulum} can be rewritten as
 \begin{equation}
 \frac{d}{dt} \, \biggl( \frac{1}{2} \, \dot{\phi}^2 -\Omega^2 \, \cos \phi \biggr) =0,
 \end{equation}
 which, with initial conditions, gives $\omega(\phi)$ ($\omega \equiv \dot{\phi}$).
 \begin{equation}
\omega^2 = \omega_o^2 +2 \Omega^2 \, (\cos \phi - \cos \phi_o).
\end{equation}
  If the system is initially stationary ($\omega_o =0$), rotational motion
 starts at the moment ablation begins --- $\Omega$ is 
 ``switched on'' --- with the angular velocity
 \begin{equation}
 \omega^2 =2 \Omega^2 \, (\cos \phi - \cos \phi_o).
\end{equation}
The angular range is restricted to $[-\phi_o, \phi_o]$ --- the dumbbell ``swings'' ---
with an amplitude-dependent period $P$, which can be expressed in terms 
of the complete elliptical integral of the first kind $K$ \citep{bel07}, according to
\begin{equation}
P=\frac{4}{\Omega} \, K \, [\sin^2 (\phi_o/2)].
\end{equation}
Using the series expansion for $K$ \citep{as72}
\begin{equation}
K[\sin^2 (\phi_o/2)] = \frac{\pi}{2} \, \biggl[ 1 + \frac{1}{4} \, \sin^2 \frac{\phi_o}{2} + O(\sin (\phi_o/2))^4 \biggr],
\end{equation}
and retaining only the first term in the limit of small $\phi_o$,
the period for a simple harmonic oscillator $2\pi/\Omega$ is recovered.
Of course, this also follows from the direct solution of Eq.\ \ref{eq:pendulum} in the
limit $\sin \phi \rightarrow \phi$.

If $\omega_o$ is non-zero, there is the possibility that the system will not reverse itself
but continue its rotational motion, albeit in a complex way compared to its initial uniform
rotation. This depends on the initial
condition pair $(\omega_o, \phi_o)$. Specifically, rotation is maintained if
\begin{equation}
\omega_o^2 > 2 \Omega^2 \, (1+\cos \phi_o).
\end{equation}
The analogy to a nonlinear pendulum still holds; in this case the initial
condition is analogous to providing a starting push to the pendulum. If its total
energy at that instant exceeds the potential energy at the top of its circular trajectory,
it will ``loop the loop,'' and continue on, absent rotational damping.

\subsection{Wedge}

We now consider two identical rectangular plates, joined to form a wedge.
The problem of a wedge is complicated by self-shadowing over a certain range of orientations
relative to the laser beam. In the interest of simplicity, we will restrict the possible range
of orientations so that self-shadowing does not occur. We use the coordinates and
set the wedge characteristics in accordance with Fig.\ 11.
The angle $\gamma$ is the half-angle separation between the two plates,
which, for simplicity, we assume to be of equal area ($A/2$). The approach here works
equally well if the plates are of unequal area, but we work with the simpler example here.
We define a length and width, $L$ and $h$, respectively, such that $A/2=Lh$. If the wedge is oriented
such that a line that bisects the angle between the wedge faces is parallel to the laser
propagation vector (along the $y$-axis), then, as we show below, this is an equilibrium position as regards
spin. Therefore, we define an angle $\phi$ to quantify angular deviations from this equilibrium.
The no-self-shadowing requirement demands only that $\phi \leq \gamma$.

\begin{figure}
\begin{center}
\includegraphics*[width=7cm,angle=0]{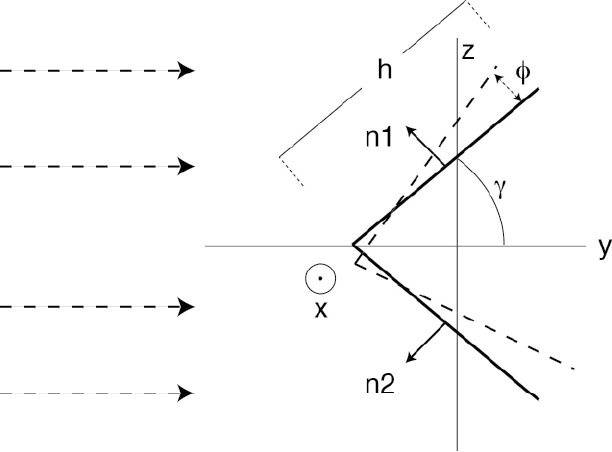}
\end{center}
\caption{Diagram of a wedge composed of two thin rectangular plates of equal area ($h \times L$;
$h$ shown, $L$, the $x$ dimension, not shown),
of arbitrary opening angle $2\gamma$ with the line of intersection parallel to the $x$-axis. 
Origin coincides with the center of mass.
Translational motion is restricted to the $yz$-plane, and rotation about the center of mass is parallel to the $x$-axis.
Laser irradiation is from the left ({\it dashed arrows}), and the
two exposed surface normals $\hat{n}_1$ and $\hat{n}_2$ are indicated. The angle $\phi$ tracks a small
angular displacement ({\it dashed lines}) from the symmetric, torque-free orientation ({\it heavy lines}).}
\end{figure}

With the two surface normals given by
\begin{equation}
\hat{n}_1=\left(
\begin{array}{c}
0 \\
-\sin(\gamma + \phi) \\
\cos (\gamma + \phi)
\end{array}
\right) ~~~~~~
\hat{n}_2=\left(
\begin{array}{c}
0 \\
-\sin(\gamma - \phi) \\
-\cos (\gamma - \phi)
\end{array}
\right),
\end{equation}
the area matrix becomes
\begin{equation}
{\bf G}=\frac{A}{2} \,
\left(
\begin{array}{ccc}
0 & 0                                                     & 0 \\
0 &  1-\cos 2\gamma \,  \cos 2\phi   & -\cos 2\gamma \, \sin 2\phi  \\
0 &  -\cos 2\gamma \, \sin 2\phi       & 1+\cos 2\gamma \,  \cos 2\phi  
\end{array}
\right),
\end{equation}
and, with $\hat{k}=\hat{y}$, the force equation for the center of mass is
\begin{equation}
m \,  \frac{d\vec{v}}{dt}=\frac{1}{2} ~ C_m I A  \, \left(
\begin{array}{c}
0 \\
1- \cos 2 \gamma \, \cos 2\phi \\
-\cos 2\gamma \, \sin 2 \phi
\end{array}
\right)
\end{equation}

\begin{figure}
\begin{center}
\includegraphics*[width=10cm,angle=0]{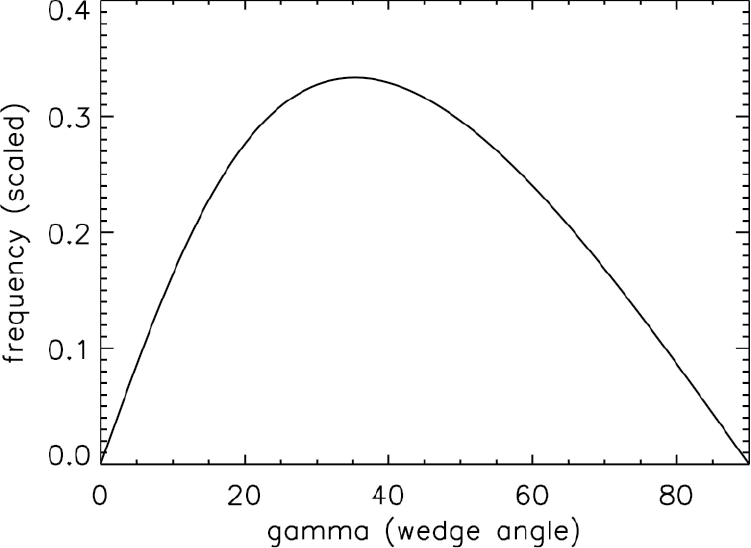}
\end{center}
\caption{Spin-dependent frequency of an irradiated wedge vs.\ the wedge half-opening angle
$\gamma$ defined by Fig.\ 11. The frequency is scaled as a multiple of $(6C_mI /\sigma h)^{1/2}$ (see text).}
\end{figure}

There are two special cases worth noting.
If $\phi=0$, there is no $z$-component to the impulse. The $y$-component becomes
\begin{equation}
m \dot{v}_y = C_mIA \,  \sin^2 \gamma,
\end{equation}
which, by symmetry, shows the expected similarity to the case of an 
arbitrarily oriented flat plate (see Eq.\  \ref{eq:flatplate}). For non-zero $\phi$, 
but $\gamma=\pi/4$ (right-angle wedge), the $z$-component of acceleration 
again vanishes, and the $y$-component becomes invariant to orientation, 
both features being analogous to the case of a cube, as discussed in \S4.1.

The moment of inertia for rotation about the center of mass, such that the angular velocity
vector is parallel to the $x$-axis, is
\begin{equation}
J_{xx}=\frac{1}{12} ~mh^2 \, (1+3 \sin^2 \gamma).
\end{equation}
We eliminate the mass in favor of a surface mass density $\sigma$ (e.g., with dimension g cm$^{-2}$),
such that $m=2 \sigma Lh$, which gives from the torque equation
%with the center of mass at $[0,(h/2)  \cos \gamma,0]$.
\begin{equation}
\ddot{\phi}=-\frac{6C_mI}{\sigma h} ~\frac{\cos^2 \gamma \, \sin^2 \gamma}{1+3 \sin^2 \gamma} ~\sin \phi,
\end{equation}
which, as was the case for an asymmetric dumbbell, is the equation of an undamped nonlinear pendulum.
In the limit of small $\phi$, the system behaves like a simple harmonic oscillator with a 
``natural'' frequency given by
\begin{equation}
\Omega^2=\frac{6C_m I}{\sigma h} ~\frac{\cos^2 \gamma \, \sin^2 \gamma}{1+3 \sin^2 \gamma},
\end{equation}
which attains a maximum value of $\Omega_{\rm max}^2 =2C_mI/3 \sigma h$, when the wedge opening 
half-angle $\gamma = \arcsin \, 3^{-1/2}$. The quantity $(\sigma h/6C_mI)^{1/2} \, \Omega$ (a dimensionless
shape-dependent frequency) is plotted against $\gamma$ in Fig.\ 12.

\subsection{Cone}

For our final example, consider a right circular cone, of uniform density $\rho$ and
mass $m$,  with opening half-angle $\alpha$,
such that $\tan \alpha = R/H$, where $R$ is the base radius and $H$ is the height, as
illustrated in Fig.\ 13. We specialize the treatment here to cases for which self-shadowing
is not present, which means that the beam vector is restricted to lie within a conical region
delineated by the dashed lines in Fig.\ 13. The more complicated case of self-shadowing
{\it can} be worked out analytically, but is somewhat tedious, so we choose to look only at the
simpler case here.

The coordinate system is shown in the lower right-hand side of Fig.\ 13, with the positive $x$-axis
pointing out of the page. If the angle $\phi$ corresponds to the azimuthal angle as defined for
standard spherical coordinate systems, then the surface normal is given by
\begin{equation}
\hat{n}=\left(
\begin{array}{c}
\cos \alpha \cos \phi  \\
\cos \alpha \sin \phi \\
\sin \alpha
\end{array}
\right).
\end{equation}
If we let $\eta \equiv H/R = \cot \alpha$, then the area matrix is
\begin{equation}
{\bf G}=\frac{\pi R^2}{(1+\eta^2)^{1/2}} \left(
\begin{array}{ccc}
\eta^2/2 &      0        & 0 \\
0             & \eta^2/2 & 0 \\
0             &      0        & 1
\end{array}
\right).
\end{equation}
We see immediately that for the special case $H=R \, \sqrt{2}$, the area matrix is
proportional
to the unit tensor, and the momentum response is precisely parallel to the beam. For all
other cases, a perpendicular component will be present.

\begin{figure}
\begin{center}
\includegraphics*[width=12cm,angle=0]{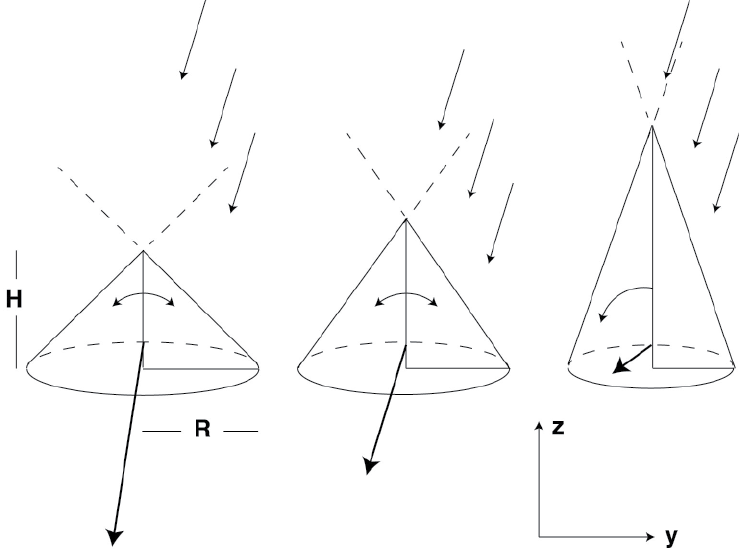}
\end{center}
\caption{Three laser-irradiated cones of equal mass with different values of $H/R$; 
1.0 ({\it left}); $\sqrt{2}$ ({\it center}); 3.0 ({\it right}). The laser propagation vector $\hat{k}$ lies
in the $yz$-plane, tilted 17$^{\circ}$ off the $z$-axis ({\it triple arrow pattern at top of each cone}).
The acceleration vectors at ablation onset are shown ({\it heavy arrows}), each lying in the $yz$-plane, with
the directions and relative magnitudes represented quantitatively. For $H/R=\sqrt{2}$ ({\it center})
the impulse is parallel to the beam, independent of the laser orientation,
while for smaller (larger) values, the impulse is nearer
(farther from) the axis. The back-and-forth arrowed arcs denote
pendulum-type oscillations, while the leftward arc on the right-side cone indicates ``tipping.'' 
For $H/R=2 \, \sqrt{2}$ (not shown), the cone is stable against torques, and does not oscillate,
also independent of the laser orientation.}
\end{figure}

The laser propagation vector is described by its angular displacement $\psi$ relative to the axis of the cone;
\begin{equation}
\hat{k}=\left(
\begin{array}{c}
0 \\
-\sin \psi \\
-\cos \psi
\end{array}
\right)
\end{equation}
so that
\begin{equation}
\hat{k} \cdot \hat{n} =-\cos \alpha  \, \sin \phi \, \sin \psi-\sin \alpha \, \cos \psi,
\end{equation}
with the restriction $\psi \leq \alpha$ satisfying the no-self-shadowing provision. We note
that the illumination condition $\hat{k} \cdot \hat{n} <0$ is, from the previous equation,
equivalent to
\begin{equation}
\sin \phi > - \tan \alpha \cot \psi,
\end{equation}
which sets the range of $\phi$-integration for the more general case.

The equation of motion becomes
\begin{equation}
\frac{d\vec{v}}{dt}=-\frac{C_mI}{(1+\eta^2)^{1/2}}  \, 
\biggl( \frac{9\pi}{\eta^2 \rho^2 m} \biggr)^{1/3}\left(
\begin{array}{c}
0 \\
(\eta^2/2) \sin \psi \\
\cos \psi
\end{array}
\right),
\end{equation}
showing that the relative parallel and perpendicular impulse components depend on both
the illumination angle and the shape of the cone.
We leave the topic of translational motion of the cone here, and move on to rotational motion.

For the following calculations, we place the origin at the center of mass.
Points on the conical surface can be specified by $z$ and $\phi$, such that
\begin{equation}
\vec{r}=\left(
\begin{array}{c}
f(z) \cos \phi \\
f(z) \sin \phi  \\
z
\end{array}
\right), 
\end{equation}
where $ f(z)=(3/4) R- (R/H) \, z$.  The domain of $z$ is $[-H/4, \, 3H/4]$,
and for the integration of Eq.\ \ref{eq:statictorque}, we have $dA=f(z) \, d\phi \, dz/\cos \alpha$.

Only the $x$-component of the cross-product in Eq.\ \ref{eq:statictorque} leads to a 
non-vanishing torque component;
\begin{equation}
\label{eq:xcross}
(\vec{r} \times \hat{n})_x=f(z) \sin \alpha \sin \phi -z \cos \alpha \sin \phi.
\end{equation}
After multiplying Eq.\ \ref{eq:xcross} by $\hat{k} \cdot \hat{n}$, and eliminating terms that vanish upon $\phi$-integration,
we are left with
\begin{equation}
N_x=-C_mI  \sin \psi \int_0^{2\pi} d\phi ~
\sin^2 \phi \int_{-H/4}^{3H/4} dz ~f(z)  ~ [f(z) \sin \alpha - z \cos \alpha ] ,
\end{equation}
which leads to
\begin{equation}
N_x= - \frac{C_m Im}{\rho} ~\frac{1-\eta^2/8}{(1+\eta^2)^{1/2}} ~\sin \psi.
\end{equation}
The torque can be positive, negative, or zero depending on the
choice of $\eta$.

\begin{figure}
\begin{center}
\includegraphics*[width=10cm,angle=0]{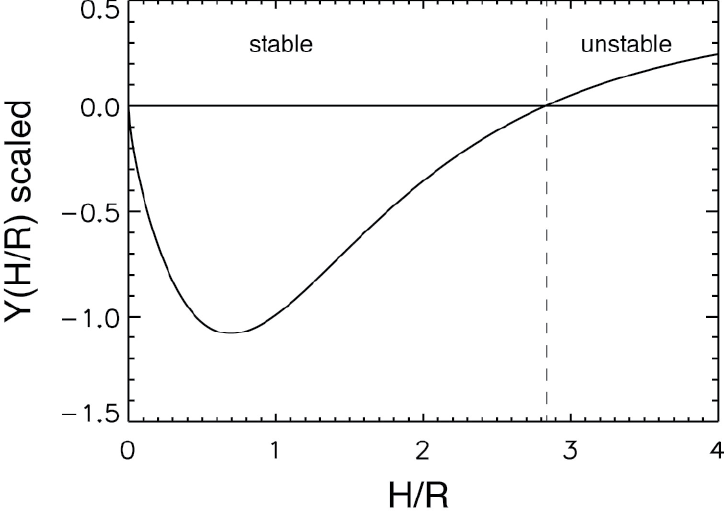}
\end{center}
\caption{
The $\eta$-dependent portion of the function $Y(\eta)$ ($\eta \equiv H/R$; see 
Eq.\ \ref{eq:yeta}), describing the rotational response of a cone to laser ablation. 
The vertical dashed line delineates regions of stable nonlinear pendulum rotation 
({\it left side}), and ``runaway tipping'' ({\it right side}).}
\end{figure}

The moment of inertia corresponding to $J_{xx}$ for rotation about the center of mass is
\begin{equation}
J_{xx}=\frac{3}{20} \, mR^2 \, \biggl(1+\frac{1}{4} \, \eta^2 \biggr)=
\frac{3}{20} \, \biggl(\frac{3}{\pi \rho \eta} \biggr)^{2/3} \, \biggl(1+\frac{1}{4} \, \eta^2 \biggr) \, m^{5/3}
\end{equation}
Therefore, after some re-arrangement, the equation of rotational motion becomes
\begin{equation}
\ddot{\psi}=Y(\eta) \,  \sin \psi,
\end{equation}
with $Y(\eta)$ given by
\begin{equation}
\label{eq:yeta}
Y(\eta)=\frac{10}{3} \, \biggl(\frac{\pi^2}{9 \rho m^2} \biggr)^{1/3} \, C_m I ~
\biggl[\frac{\eta^{2/3}}{\sqrt{1+\eta^2}} \, \frac{\eta^2 -8}{\eta^2 +4} \biggr].
\end{equation}
Since $Y$ can be positive, negative, or zero, there are three types of rotational behavior, 
pendulum-type oscillation for cones with $H<2 R \sqrt{2}$, unstable ``tipping'' for $H> 2 R \sqrt{2}$,
and metastability against rotation for $H=2 R \sqrt{2}$. The bracketed factor in Eq.\ \ref{eq:yeta} is
plotted in Fig.\  14.

\section{Discussion}

Our goals here are: (1) to provide the means by which to adjust or augment some of the intuitive notions
that have been built upon one-dimensional modeling and experiments that involve
flat plates, and (2) to provide a framework for carrying out more elaborate orbital calculations
than what are required under the assumption that potential debris targets are spheres or ideally aligned flat plates. 
Still, even beyond technological questions related to our ability to accurately track and engage small debris
from a ground station [see discussion in \citet{phipps2012} and references therein], a capability which is pre-supposed here,
there are a number of additional refinements that 
are likely necessary to further quantify the response of an object to laser ablation,
as discussed in the remainder of this section.

We can look at laser-induced shape deformation in two ways: (1) smooth, surface-wide deformation, and
(2) creation of local smaller scale irregularities. The first category might be exemplified by the problem
treated in \S5.3, a rotating sphere subjected to ablation that is inherently non-uniform, owing simply to
the curvature of the surface. The second category
includes intrinsic surface irregularities that change the local intensity on target relative to the perfect surfaces
we have assumed (a $\hat{k} \cdot \hat{n}$ effect), 
as well as irregularities that may be created by beam imperfections, particularly those
caused by atmospheric propagation. Although adaptive optics techniques mitigate this latter effect, deviations from
perfection are not unlikely. Spatial beam inhomogeneities result in variations in the local mass loss rate,
which leads to non-uniform surface deformation. Moreover, the presence of local variations in the incident intensity means
that, strictly speaking, the coupling itself varies locally, irrespective of surface irregularities. But it is worth noting that if
small-scale irregularities tend to reduce the incident energy deposition rate via the $\hat{k} \cdot \hat{n}$ dependence,
such reductions tend to improve the coupling for the operating range of interest, partially offsetting
the pure intensity dependence of the local mass loss rate. Progressive surface deformation and its effect
on momentum transfer could be included in more exacting numerical work. No doubt, laboratory 
measurements of the evolution of a target subjected to multiple ablation events would prove useful.

We noted in \S2.2 our assumption that $C_m$ is treated in this paper as a constant across the illuminated
surface of a given object. While
we argued for the validity of this approximation, a suggested improvement for future work is to
account explicitly for the intensity dependence of the coupling: $C_m \rightarrow C_m(I)$. Again, the
sparseness of the experimental database becomes a factor in accurately invoking this functional dependence.
We stress that the quantities of primary interest to debris clearing 
are directly proportional to the magnitude of the coupling coefficient
(recoil velocities, spin damping constants), or to its square root (induced spin frequencies).
Accurate numerical predictions of momentum transfer presuppose an adequately comprehensive 
and experimentally precise database of coupling coefficients. 

A potential concern to debris clearing programs can be expressed in terms of the well-known maxim
of medical ethics: ``first, do no harm''; one must ask whether or not laser ablation will simply
exacerbate the problem by creating many fragments from one. 
The target recoil during an ablation event creates a pressure pulse in the target material. 
Experimental data can be approximated by the expression \citep{phipps88,fabbro}
\begin{equation}
\label{eq:pressure}
P=3.9 \, I^{ \, 0.7} \, \lambda^{-0.3} \, \tau^{-0.15} ~~{\rm kbars},
\end{equation}
where the intensity, laser wavelength, and pulse duration are expressed as multiples of
GW cm$^{-2}$, $\mu$m, and ns, respectively. (Some care with Eq.\ \ref{eq:pressure} should
be exercised, since it describes a trend line only.) The response of a debris object to a pressure pulse is highly
material-dependent. It is not unlikely that some spallation may occur \citep{zeldovich}. 
The spall thickness --- the characteristic size of potential spallation fragments --- is roughly the 
material sound speed times the pulse duration. For sound speeds of a few times $10^5$ cm s$^{-1}$, 
and a pulse duration of a few nanoseconds, the spall thickness $\sim10 ~ \mu$.
Although these spallation products would not pose a direct threat to space assets, the subsequent 
laser/target momentum coupling may be affected by the resulting shape modification. Laboratory 
studies in the ns-pulse $10^8$--$10^{10}$ W cm$^{-2}$ regime of progressive target deformation 
caused by the combination of ablation and possible spallation events are desirable in order to identify 
the prevalence and overall nature of this type of shape evolution. 

Finally, we note that our goal (2) above is only partially addressed here. We have said nothing about the
consequences for calculations of orbital modifications, which are essential to planning sustained laser engagements,
target re-acquisition when multiple-orbit engagements are needed, and ultimately, optimizing
the perigee reduction per unit laser energy expended. We have shown that targets will typically
acquire off-beam impulse components. Calculations of orbital angular momentum transfer and energy transfer
either to or from a debris fragment must include a new set of vectors, namely, those that describe the illuminated
target surfaces.
In terms of engaging spinning targets, we have shown that assuming a net impulse along the beam
by spin-averaging does not suffice to predict resultant orbits, and that off-beam deflections depend
on the spin state of the target at the onset of ablation. Also, target spin
cannot be de-coupled from the linear momentum transfer calculations, since the area matrix becomes
time-dependent in that case. In that sense, the efficiency of linear momentum transfer is also dependent upon
the initial spin state. In future work, we will begin to examine the consequent effects on orbital trajectories in
low-Earth orbit.

The relations among the laser station coordinates, orbital parameters, and selection
of laser engagement time intervals are made more complex by having to deal with shape effects. However, these
complications are manageable, as shown by our analyses. In fact, they can be used to help refine predictive
simulations of debris de-orbiting campaigns, enabling more precise descriptions of the behavior of the debris 
population in response to laser ablation. While the calculations in this work have been carried out for a few 
idealized shapes, our methods can be used with 3-D models of real objects, invoking a variety of materials 
and shapes, to develop a comprehensive modeling package for pulsed laser de-orbiting or deflection of space debris.

\section*{Acknowledgements}
The authors thank the referees for providing useful suggestions that have helped to improve the clarity of 
the manuscript. Lawrence Livermore National Laboratory is operated by Lawrence Livermore National 
Security, LLC, for the U.S. Department of Energy, National Nuclear Security Administration under 
Contract DE-AC52-07NA27344.

%\section*{References}

%%%%%%%%%%%%%%%%%%%%%%%%%%%%%%%%%%%%%%%%%%%%%%%%%%%%%%%%%%%%%%%%%%%%%%%%%%%%%
%% Appendices
% \appendix

%%%%%%%%%%%%%%%%%%%%%%%%%%%%%%%%%%%%%%%%%%%%%%%%%%%%%%%%%%%%%%%%%%%%%%%%%%%%%
%% Appendices
%\appendix{appendixheading}

\end{document}